\documentstyle[preprint,aps,floats,epsf]{revtex}

\begin{document}

\newcommand{\s}{\rm s}
\newcommand{\id}{\rm ID}
\newcommand{\cm}{\rm cm}
\newcommand{\km}{\rm km}
\newcommand{\gev}{\rm GeV}
\newcommand{\tev}{\rm TeV}
\newcommand{\susy}{\rm SUSY}

\tightenlines 


\preprint{\font\fortssbx=cmssbx10 scaled \magstep2
\hbox to \hsize{
\hfill$\vcenter{\hbox{\bf MADPH-00-1195}
		\hbox{\bf OKHEP-00-08}
                \hbox{\bf hep-ph/0105182}
                \hbox{May 2001}}$ }
}

\title{\vspace{.5in}
Indirect Search for Neutralino Dark Matter \\
with High Energy Neutrinos}

\author{
V. Barger$^a$, Francis Halzen$^a$, Dan Hooper$^a$ and Chung Kao$^b$}

\address{
$^a$Department of Physics, University of Wisconsin,
Madison, WI 53706 \\
$^b$Department of Physics and Astronomy, University of Oklahoma,
Norman,  OK 73019
}

\maketitle

\bigskip

\begin{abstract}

We investigate the prospects of indirect searches for supersymmetric
neutralino dark matter. Relic neutralinos gravitationally accumulate
in the Sun and their annihilations produce high energy
neutrinos. Muon neutrinos of this origin can be seen in large detectors
like AMANDA, IceCube and ANTARES.
We evaluate the relic density and the detection rate in several models
--- the minimal supersymmetric model, minimal supergravity,
and supergravity with non-universal Higgs boson masses
at the grand unification scale.
We make realistic estimates for the indirect detection rates  
including effects of the muon detection threshold, quark hadronization,
and solar absorption.
We find good prospects for detection of neutralinos with mass above 200 GeV.
\end{abstract}

\pacs{PACS numbers: 12.60.Jv, 13.85Qk, 14.60.Lm, 14.80.Ly}
%

\newpage

\section{Introduction}

Average matter and energy densities in the Universe ($\rho_i$)
are commonly described in terms of density parameters ($\Omega_i =  
\rho_i/\rho_c$), where
\begin{eqnarray}
\rho_c = 3H_0^2/(8\pi G_N) \simeq 1.88 \times 10^{-29} h^2 \;
{\rm g/cm^3},
\end{eqnarray}
is the critical density to close the Universe,
$h = H_0/( 100 \; {\km}\; {\rm sec}^{-1}\; {\rm Mpc}^{-1})$,
$H_0$ is the Hubble constant,
and $G_N$ is Newton's gravitational constant.
Recent measurements of the Hubble constant
are converging to $h \simeq 0.6 - 0.7$ \cite{Hubble1,Hubble2,Hubble3}.

Studies of clusters of galaxies and large scale structure
indicate that the matter density should be
at least $\Omega_M \agt 0.2$ \cite{Rotation}.
The matter density inferred from observations of
cluster X-ray \cite{Xray}, supernovae \cite{Supernova},
and the cosmic microwave background radiation (CMB)
anisotropy data \cite{BOOMERANG,MAXIMA,DASI} 
is $\Omega_M \simeq 0.3-0.4$.
The baryon density inferred from Big-Bang Nucleosynthesis is
$\Omega_b h^2 \simeq 0.020\pm 0.002$ \cite{Baryon} 
which is in good agreement with the value obtained from recent analyses of
the CMB power spectrum \cite{BOOMERANG,DASI}.
The luminous matter in the Universe has a very small relative density
$\Omega_{\rm L} \alt 0.01$ \cite{visible}.
Therefore, the evidence
for substantial dark matter is compelling.
Moreover, the dominant component must be cold dark matter
to be consistent with large scale structure \cite{cdm-lss}.
The most attractive cold dark matter candidates
are stable weakly interacting mass particles (WIMPs)
produced in the early Universe.
In our analysis, we take $\Omega_{\chi} = \Omega_M -\Omega_b$ 
to be the WIMP density and conservatively consider
\begin{equation}
0.05 \alt \Omega_{\chi} h^2 \alt 0.3.
\label{eq:ohs}
\end{equation}
as the cosmologically interesting region for the cold dark matter density .

In the early Universe, the temperature ($T$) was much higher than
the WIMP mass ($m_\chi$) and WIMPs existed abundantly
in thermal equilibrium with WIMP annihilation into lighter particles
balanced by pair production. When the Universe cooled
and the temperature fell below $m_\chi$, the WIMP density
became suppressed by the Boltzmann factor exp($-m_\chi/T$).
As a result of the expansion of the Universe, 
WIMPs dropped out of equilibrium  
with other particles at some point, leaving a relic abundance.

{}From simple dimensional analysis, the WIMP annihilation cross section
is about
\begin{eqnarray}
\sigma \sim \alpha^2/m_\chi^2,
\end{eqnarray}
and their relative density is
\begin{eqnarray}
\Omega_\chi h^2 = m_\chi n_\chi/\rho_c
               \simeq (3 \times 10^{-27} \; \cm^3/s)/<\sigma v>
               \sim (3 \times 10^{-27} \; \cm^3/s) (m_\chi^2/ \alpha^2),
\end{eqnarray}
where $\alpha$ is the fine-structure constant.
For a WIMP mass of the order of the weak boson mass, $m_\chi \sim m_W$,
the WIMP density is $\Omega_\chi \sim 1$.
Thus, a cosmologically interesting $\Omega_\chi$ is natural in theories
with a weak scale WIMP mass.

In supersymmetric (SUSY) theories with a conserved
$R$-parity\footnote{$R = +1$ for particles in the Standard Model
as well as Higgs bosons and $R = -1$ for their superpartners.},
the lightest SUSY particle (LSP) is stable and is
thus an attractive WIMP candidate 
\cite{SUSYDM1,SUSYDM2,Jungman1996}.
Most commonly, the LSP is the lightest neutralino which is
a superposition of the supersymmetric partners of the photon,
the $Z$ boson and the neutral Higgs bosons.
The gaugino and higgsino composition of the neutralino is
an important factor in the neutralino annihilation cross section.
If the neutralinos are more gaugino-like,
they annihilate mostly into $b\bar{b}$ and $\tau\bar{\tau}$;
if the neutralinos are higgsino-like,
most of them annihilate into gauge bosons and $t\bar{t}$.

WIMPs in the halo of our galaxy
lose energy when they pass through the Sun and scatter from nuclei.
When their velocities fall below the escape velocity from the Sun, 
they become gravitationally trapped and accumulate.
Annihilations of these WIMPs are a potential source of high energy
neutrinos that can be detected via Cerenkov light by large under-ice
or under-water photomultiplier arrays in neutrino telescopes,
such as AMANDA\cite{amanda}, IceCube\cite{icecube} 
and ANTARES\cite{antares}.

In this paper we investigate the prospects of observing
muon neutrinos produced by WIMP annihilation in the Sun.
Due to cosmic ray backgrounds, only the signals from the muon neutrinos 
at or below the horizontal detection can be detected.
In Section II, we estimate the general expectations
for neutrino signals from WIMP annihilation,
the effect of the muon detection energy threshold, and
the backgrounds to the signal.
In Sections III, IV and V, we evaluate the indirect detection rate
within the framework of the minimal supersymmetric model (MSSM),
the minimal supergravity model (mSUGRA),
and supergravity unified models with non-universal Higgs boson masses
at the grand unified scale, respectively.
In Section VI and VII, we summarize the prospects of indirect search 
and our conclusions about indirect detection of neutralino dark matter in large ice
or water experiments.

\section{Indirect Detection of WIMPs}

When our galaxy formed, the cold dark matter WIMPs were clustered
with the luminous matter and formed a significant fraction
of the galactic matter density \cite{density1,density2}
\begin{eqnarray}
\rho_\chi \simeq 0.3-0.5 \; {\gev/\cm^3}
\end{eqnarray}
as deduced from rotation curves.
The distribution of WIMPs in the galactic halo is usually approximated
as an isothermal sphere with an average velocity
\cite{velocity1,velocity2,velocity3}
\begin{eqnarray}
 v_\chi \simeq 220-300 \; {\km/sec}.
\end{eqnarray}
The density of WIMPs that are gravitationally trapped in the Sun
builds up until equilibrium is reached when their annihilation rate
is half of the capture rate (the factor of 2 is due to the Majorana nature of WIMPs).
Since the WIMP mass is naturally in the 100 GeV--1000 GeV mass range,
they annihilate mostly into weak bosons and heavy quarks, which
subsequently decay into final states with high-energy neutrinos
\cite{Jungman1996,Halzen1995}.
The high-energy muon neutrinos can be detected in large ice or water detectors
by the Cherenkov light from the muons.
The muons in these events are aligned along a direction pointing back to the Sun.
The angular spread of the muons from this direction is
$\theta_\mu \sim 1.2^o/\sqrt{E_\mu (\tev)}$.
The atmospheric neutrino events in a pixel containing
this signal are the backgrounds. 
The measurement of muon energy might be used to
(i) focus the search window around the direction from the Sun
in order to reduce the background and
(ii) estimate the neutrino energy
from the angular spread.

\subsection{Indirect Detection Rate}

In this subsection, we make a rough estimate of the indirect detection rate
for a generic WIMP dark matter under the following assumptions\footnote{
In this section
we take $\rho_\chi = 0.4 \; \gev/\rm cm^3$ and $v_\chi = 300 \; \km/s$.}
\cite{Halzen1995}:

\begin{enumerate}
\item[(i)] the measured galactic halo density is mostly associated with WIMPs
and their flux is
\begin{eqnarray}
\phi_\chi
 & = & n_\chi v_\chi
       \simeq \frac{0.4 \; \cm^{-3}}{m_\chi\rm\ (GeV)} 
              \times (3\times 10^7 \; {\rm cm}/{\rm s}) \nonumber \\
 & \simeq & \frac{1.2\times 10^7}{m_\chi\;(\gev)} {\cm^{-2}}{\rm s^{-1}},
\end{eqnarray}
where $m_\chi$ is the WIMP mass;

\item[(ii)] the WIMP-nucleon cross section is approximately given by
the dimensional analysis result
\begin{eqnarray}
\sigma(\chi N) \equiv \sigma_{\rm DA}
\simeq (G_F m_N^2)^2 \frac{1}{m_W^2} = 6 \times 10^{-42} \; \cm^2;
\end{eqnarray}

\item[(iii)] the WIMPs annihilate about 10\% of the time into neutrinos
via the channels $\chi\bar{\chi} \to W^+W^-$ or $Q\bar{Q}$
where $Q$ is a heavy quark (bottom or top).
\end{enumerate}

The further physical quantities needed to calculate
the indirect detection rate for neutrinos from WIMP annihilation
in the Sun are as follows:

\begin{enumerate}
\item[(i)] the capture cross section in the Sun ($\sigma_\odot$),
given by the product of the number of target nucleons in the Sun and
the elastic $\chi N$ cross section ($\sigma_{El}$) and 
a focusing  factor ($f \simeq 10$) that is given by the ratio of kinetic
and potential energy of the WIMP near the Sun \cite{Gaisser1995}.
\begin{equation}
\sigma_\odot = f  \left[ 1.2\times10^{57} \right] \sigma_{\rm El} \;.
\label{sigma Sun}
\end{equation}

\item[(ii)] the neutrino flux at the Earth from WIMP annihilation in the Sun
\begin{equation}
\phi_\nu = B_\chi \phi_\chi \sigma_\odot / 4\pi d^2 \;,
\label{phi Sun}
\end{equation}
where $d = 1 \, {\rm A.U.} = 1.5\times10^{13}\rm \,cm$ and $B_\chi$ is the  
branching fraction to neutrinos ($\sim 10\%$). Thus we find
\begin{equation}
\phi_\nu = {3\times10^{-5}\over m_{\chi}\rm\ (GeV)} \, \cm^{-2} \, \s^{-1} \;.
\end{equation}

\item[(iii)] the probability of neutrino detection \cite{Gaisser1995}, given by
\begin{eqnarray}
P = N_A \sigma_\nu R_\mu \simeq 2\times 10^{-13} \, \left[ m_{\chi}\rm\ (GeV)  
\right]^2,
\end{eqnarray}
where $N_A \simeq 6\times10^{23} $ is the Avogadro number,
$\sigma_\nu = 0.5\times10^{-38} E_\nu \, (\gev) \ \cm^{2}$ is the average of
the $\nu N$ and  $\bar{\nu} N$ cross sections, and
$R_\mu \simeq 500\, \cm \ \times E_\mu \, (\gev)$ is the muon range.
For the kinematics of the annihilation and decay chain
\begin{eqnarray}
\chi\bar\chi &\to& W^+W^- \nonumber \\  \noalign{\vskip-1ex}
             &   & \hspace{2em} \raise1ex\hbox{$\vert$}\!{\rightarrow}
\mu\nu_\mu \nonumber
\end{eqnarray}
the average neutrino energy is about $<E_\nu> \simeq {1\over2}m_\chi$. 
For the decay chain into heavy quarks 
\begin{eqnarray}
\chi\bar\chi &\to& Q\bar{Q} \nonumber \\  \noalign{\vskip-1ex}
             &   & \hspace{2em} \raise1ex\hbox{$\vert$}\!{\rightarrow}
q\mu\nu_\mu \nonumber
\end{eqnarray}
the average neutrino energy is about $<E_\nu> \simeq {1\over3}m_\chi$. 

\item[(iv)] the number of neutrino events per year per unit area (m$^2$)
detectable in a neutrino telescope from $\chi\bar{\chi}$ annihilation
in the Sun 
\begin{eqnarray}
dN_{\id}/dA = \phi_{\nu} P \simeq 2 \times10^{-6} \,
m_{\chi}  \rm\ (GeV)\ (year)^{-1} \, (m^2)^{-1}.
\end{eqnarray}
\end{enumerate}

For an effective area of $10^4$~m$^2$, the indirect detection rate is approximately
\begin{eqnarray}
dN_{\rm ID}/ dA \simeq 2 \times10^{-2} m_{\chi} \rm\ (GeV)\ (10^4\,m^2)^{-1}  
(year)^{-1} \,.
\end{eqnarray}

\subsection{Neutrino-generated Muons}

Neutrino telescopes detect muons from the charged-current reaction 
\begin{equation}
\nu_\mu + N \to \mu +X,
\end{equation}
where $N$ is an average nucleon in the Earth below the detector.
The effective detection volume is the product of
the detector area and the muon range in rock ($R_\mu$)
which increases with energy.
By observing muons produced by charged current of
$\nu_\mu(\bar{\nu}_\mu)$ interactions in the rock below
the detector, the effective volume is enhanced.
The characteristic range of TeV muons is about one kilometer
in rock, so large effective volumes are realized
at TeV energies.
This technique applies only for upward going muons entering
the detector from below the horizontal,
because of the large backgrounds from atmospheric muons
that dominate neutrino signals from above the horizontal.

The average muon energy loss rate is
\begin{equation}
{\left\langle{\rm d}E\over{\rm d}X\right\rangle}\;=\;
-\alpha(E)\,-\,\beta(E)\times E,
\label{muloss}
\end{equation}
where $X$ is the thickness of material in g/cm$^2$.
The first term represents ionization losses, which are
approximately continuous, with $\alpha\sim 2$~MeV~g$^{-1}$cm$^2$.
The second term includes the catastrophic processes of bremsstrahlung,
pair production and nuclear interactions, in which fluctuations
play an essential role.
Here $\beta\sim 4\times 10^{-6}$~g$^{-1}$cm$^2$.
The critical energy above which the radiative processes
dominate is
\begin{equation}
E_{\rm cr}\;=\;\alpha/\beta\;\approx\;500\;{\gev}.
\end{equation}

To treat muon propagation properly when $E_\mu>E_{\rm cr}$ requires
a Monte Carlo calculation of the probability $P_{\rm surv}$ that
a muon of energy $E_\mu$ survives with energy $>E_\mu^{\rm min}$
after propagating a distance $X$ \cite{Lipari}.
The probability that a neutrino of energy $E_\nu$
on a trajectory through a detector produces a muon above
the threshold energy at the detector is
\cite{Lipari,Gaisser1985,Gaisser1987}
\begin{equation}
P_\nu(E_\nu,E_\mu^{\rm min})\,=\,N_A\,\int_0^{E_\nu}\,{\rm d}
E_\mu{{\rm d}\sigma_\nu\over{\rm d}E_\mu}(E_\mu,E_\nu)\,
R_{\rm eff}(E_\mu,E_\mu^{\rm min}),
\label{P_nu E}
\end{equation}
where
\begin{equation}
R_{\rm eff}\,=\,\int_0^\infty
\,{\rm d}X\,P_{\rm surv}(E_\mu,E_\mu^{\rm min},X).
\end{equation}
The flux of $\nu_\mu$-generated muons at the detector
is given by a convolution of the neutrino spectrum $\phi_\nu$
with the muon production probability (\ref{P_nu E}) as
\begin{equation}
\phi_\mu(E_\mu^{\rm min},\theta)\;=\;
\int_{E_\mu^{\rm min}}\, {\rm d}E_\nu \, P_\nu(E_\nu,E_\mu^{\rm min})\,
\exp[-\sigma_{\rm tot}(E_\nu)\,N_A\,X(\theta)]\,\phi_\nu(E_\nu,\theta).
\label{N_mu}
\end{equation}
The exponential factor in the integrand accounts for absorption
of neutrinos along the chord through the Earth, $X(\theta)$.
Absorption becomes important for $\sigma(E_\nu)\agt 10^{-33}$~cm$^2$
or $E_\nu\agt 10^7$~GeV.
The event rate is calculated by multiplying Eq.~(\ref{N_mu})
with the effective area of the detector.

The probability $P_\nu(E_\nu,0)$ can be approximated with a power law
\begin{equation}
 P_{\nu\to\mu} \simeq 1.3\times10^{-6}\,(\frac{E}{\rm TeV})^{2.2}
\end{equation}
for $E_\nu$ between 1 GeV and 1 TeV.
This energy dependence reflects the neutrino cross section
($\sigma_\nu \propto E$) and
the effective muon range ($R_\mu \propto E$) in Eq.~(\ref{P_nu E}).
Above 1~TeV, the effect of the $W$ propagator slows the growth
of $\sigma_\nu$ and the muon range also makes a transition
from linearity to a more gradual energy dependence.

\subsection{Atmospheric Neutrino Backgrounds}

The background to indirect detection (ID) is determined by the flux
of atmospheric neutrinos in a pixel region of the detector that
subtends the Sun. 
The number of upward atmospheric background events per year
in a $10^4$~m$^2$ detector is $\sim 10^2/E_\mu$(TeV).
The size of the pixel region is determined by the angle between muon and
neutrino ($\sim 1.2^\circ \Big/ \sqrt{E_\mu(\rm TeV)}$).
Using the kinematics $E_\mu \simeq m_\chi/4$ for the WIMP signal from  
annihilation to $W$-pairs, we obtain
\[
B_{\rm ID} = { 10^2/E_\mu({\rm TeV}) \over 2\pi \Big/
\left[ 1.2^\circ {\pi\over 180^\circ} \over \sqrt{E_\mu(\rm TeV)}
\right]^2}
= {1.1\times10^5\over m_{\chi}^2\rm\ (GeV)^2}
\mbox{10$^4$\,m$^2$}^{-1} \, {\rm year}^{-1}.
\]
This expression is only valid for for $E_\mu \approx m_\chi/4 >100$~GeV.
More generally, the backgrounds are given in Table I.
\begin{table}[h]
\centering\unskip
\caption{Number of background events due to atmospheric neutrinos.}
\tabcolsep=1.25em
\begin{tabular}{c|ccc}
\hline
 &\small N(Background) &\small Number of pixels &\small N(Background)
\\
\noalign{\vskip-.85ex}
 &\small in 10$^4$\,m$^2$ &\small of solar size  &\small per 10$^4$\,m$^2$\\
\noalign{\vskip-.85ex}
$E_\mu$(GeV) &\small    &\small in 2$\pi$ &\small per pixel, per year \\
\hline
10& 3200& 140& 23\\
100& 1060& $1.4\times10^3$& 0.8\\
1000& 110& $1.4\times10^4$& $8\times10^{-3}$\\
\hline
\end{tabular}
\end{table}

For large $m_\chi$ the signal to background ratio is
\[
\left(N\over B\right)_{\rm ID} \equiv {dN_{\rm ID}/dA\over dB_{\rm ID}/dA}
\simeq 7.2\times10^{-6} m_{\chi}^3\rm\ (GeV)^3 \,.
\]
Table II shows the expected signal and signal to background estimates
for several values of $m_\chi$. Thus high-energy muons pointing at the Sun  
are a viable signal for sufficiently high $m_\chi$.

\begin{table}[h]
\caption{Event rates and signal to background $(N_S/N_B)$ ratio
for an effective detector area of $10^4$ m$^2$ in a year.}
\centering\unskip\smallskip
\tabcolsep=1.5em
\begin{tabular}{c|c@{\quad}cc@{\quad}c}
\hline
$m_\chi$ (GeV) & Indirect detection rate (events/$10^4$\,m$^2$/year)
& $N_S/N_B$ \\
\hline
50   & $2.3\times10^1$  & $\simeq\,1$ \\
500  & $2\times10^2$    & $\simeq\,10^2$ \\
2000 & $1.7\times10^2$  & $\simeq\,10^4$ \\
\hline
\end{tabular}
\end{table}

In the next three sections we make quantitative calculations of the WIMP
annihilation rates expected in supersymmetry and supergravity models 
to replace the simple estimate from Eq.(8).

\section{The Minimal Supersymmetric Model}

A supersymmetry (SUSY) between fermions and bosons is a
compelling extension of the Standard Model (SM) for several reasons.

(i) It provides an elegant solution to the fine tuning problem;

(ii) With the particle mass spectra of the minimal supersymmetric
standard model (MSSM) the evolution of renormalization group equations
is consistent with a grand unified scale
$M_{\rm GUT} \sim 2\times 10^{16}$ GeV
and a SUSY mass scale in the range $M_Z \alt M_{\rm SUSY} \alt 1$ TeV;

(iii) With a large top quark Yukawa coupling at a grand unified scale
($M_{\rm GUT}$),  radiative corrections can drive a Higgs boson mass
squared parameter negative,
spontaneously breaking the electroweak symmetry,
and naturally explain the origin of the electroweak scale;

(iv) If $R$-parity\footnote{
The $R-$parity is defined as $R = (-1)^{3B+L+2S}$,
where $B$ = Baryon number, $L$ = Lepton number and $S$= Spin.
$R = +1$ for the SM particles and Higgs bosons,
and $R = -1$ for their superpartners.}
is conserved, the lightest neutralino in supersymmetric models
can be a candidate for cold dark matter;

The minimal supersymmetric standard model (MSSM) \cite{MSSM}
is the minimal extension of the standard model (SM) with
(i) a supersymmetric partner for each SM particle and
every Higgs boson,
(ii) two Higgs doublets $H_1$ and $H_2$ such that
$H_1$ couples to the fermions with $t_3 = -1/2$ and
$H_2$ couples to the fermions with $t_3 = +1/2$,
(iii) a Higgs mixing parameter $\mu$ assumed to be real,
(iv) three real symmetric $3\times 3$ matrices
      of squark mass-squared parameters,
      $M_{\tilde{Q}}^2$, $M_{\tilde{U}}^2$ and $M_{\tilde{D}}^2$,
(v) two real symmetric $3\times 3$ matrices
      of slepton mass-squared parameters,
      $M_{\tilde{L}}^2$ and $M_{\tilde{E}}^2$,
(vi) three real $3\times 3$ matrices of trilinear Yukawa parameters,
      $A_U$, $A_D$ and $A_E$, and
(vii) a conserved $R$-parity
$R \equiv (-1)^{3B+L+2S} = (-1)^{3(B-L)+2S}$.
In the MSSM, the lightest supersymmetric particle (LSP) is stable
and is a very attractive candidate
for cosmological dark matter.

The neutral electroweak gauginos ($\tilde{B}$ and $\tilde{W}^3$)
and the Higgsinos ($\tilde{H}^0_1$ and $\tilde{H}^0_2$)
have the same quantum numbers.
They mix and generate four mass eigenstates called neutralinos.
The neutralino mass matrix in the basis of
($\tilde{B},\tilde{W}^3,\tilde{H}^0_1,\tilde{H}^0_2$) states is
\begin{equation}
\arraycolsep=0.01in
{\cal M}_N=\left( \begin{array}{cccc}
M_1 & 0 & -M_Z\cos \beta \sin \theta_W^{} & M_Z\sin \beta \sin \theta_W^{}
\\
0 & M_2 & M_Z\cos \beta \cos \theta_W^{} & -M_Z\sin \beta \cos \theta_W^{}
\\
-M_Z\cos \beta \sin \theta_W^{} & M_Z\cos \beta \cos \theta_W^{} & 0 & -\mu
\\
M_Z\sin \beta \sin \theta_W^{} & -M_Z\sin \beta \cos \theta_W^{} & -\mu & 0
\end{array} \right)\;.
\end{equation}
This mass matrix is symmetric and can be diagonalized
by a single matrix \cite{MSSM}.
The diagonalization is given by the following analytic
expressions \cite{ElKheishen,BBO}
\begin{mathletters}
\begin{eqnarray}
\epsilon _1M_{\chi _1^0}&=&-({1\over 2}a-{1\over 6}C_2)^{1/2}
+\left [-{1\over 2}a-{1\over 3}C_2+{{C_3}\over {(8a-{8\over 3}C_2)^{1/2}}}
\right ]^{1/2}+{1\over 4}(M_1+M_2)\;, \nonumber \\ \\
\epsilon _2M_{\chi _2^0}&=&+({1\over 2}a-{1\over 6}C_2)^{1/2}
-\left [-{1\over 2}a-{1\over 3}C_2-{{C_3}\over {(8a-{8\over 3}C_2)^{1/2}}}
\right ]^{1/2}+{1\over 4}(M_1+M_2)\;, \nonumber \\ \\
\epsilon _3M_{\chi _3^0}&=&-({1\over 2}a-{1\over 6}C_2)^{1/2}
-\left [-{1\over 2}a-{1\over 3}C_2+{{C_3}\over {(8a-{8\over 3}C_2)^{1/2}}}
\right ]^{1/2}+{1\over 4}(M_1+M_2)\;, \nonumber \\ \\
\epsilon _4M_{\chi _4^0}&=&+({1\over 2}a-{1\over 6}C_2)^{1/2}
+\left [-{1\over 2}a-{1\over 3}C_2-{{C_3}\over {(8a-{8\over 3}C_2)^{1/2}}}
\right ]^{1/2}+{1\over 4}(M_1+M_2)\;, \nonumber \\
\end{eqnarray}
\end{mathletters}
where $\epsilon _i$ is the sign of the $i$th eigenvalue
of the neutralino mass matrix, and
\begin{mathletters}
\begin{eqnarray}
C_2&=&(M_1M_2-M_Z^2-\mu^2)-{3\over 8}(M_1+M_2)^2\;, \\
C_3&=&-{1\over 8}(M_1+M_2)^3
+{1\over 2}(M_1+M_2)(M_1M_2-M_Z^2-\mu^2)+(M_1+M_2)\mu^2
\nonumber \\
&&+(M_1\cos ^2\theta _W^{}
+M_2\sin ^2\theta _W^{})M_Z^2 -\mu M_Z^2\sin 2\beta\;, \\
C_4&=& (M_1\cos ^2\theta _W^{}+M_2\sin ^2\theta _W^{})
M_Z^2\mu\sin 2\beta -M_1M_2\mu^2\nonumber \\
&&+{1\over 4}(M_1+M_2)
[(M_1+M_2)\mu^2+(M_1\cos ^2\theta _W^{}+M_2\sin ^2\theta _W^{}) M_Z^2
-\mu M_Z^2\sin 2\beta] \nonumber \\
&&+{1\over 16}(M_1M_2-M_Z^2-\mu^2)(M_1+M_2)^2
-{3\over 256}(M_1+M_2)^4\;, \\
a&=&{1\over {2^{1/3}}}{\rm Re}\left [-S+i(D/27)^{1/2}\right ]^{1/3}\;, \\
D&=&-4U^3-27S^2\;, \quad U=-{1\over 3}C_2^2-4C_4, \quad S=-C_3^2
-{2\over 27}C_2^3  +{8\over 3}C_2C_4\;.
\end{eqnarray}
\end{mathletters}
The masses given by the above expressions are usually not
in the order
$M_{\chi_1^0}<M_{\chi_2^0}<M_{\chi_3^0}<M_{\chi_4^0}$,
but the eigenstates can be relabeled.

The mixing matrix $Z$, 
defined by ${\cal M}_{\rm diag} = Z {\cal M}_N Z^{-1}$, is given by  
analytic expressions
\cite{ElKheishen,BBO}
\begin{mathletters}
\begin{eqnarray}
{{z_{i2}}\over {z_{i1}}}&=&-{1\over {\tan \theta _W^{}}}
{{M_1-\epsilon _iM_{\chi _i^0}}\over {M_2-\epsilon _iM_{\chi _i^0}}}\;,
\\
{{z_{i3}}\over {z_{i1}}}&=&{{ \mu [M_2-\epsilon _iM_{\chi _i^0}]
[M_1-\epsilon _iM_{\chi _i^0}]-M_Z^2\sin \beta \cos \beta
[(M_1-M_2)\cos ^2\theta _W^{}+M_2-\epsilon _iM_{\chi _i^0}]}\over
{M_Z[M_2-\epsilon _iM_{\chi _i^0}]\sin \theta _W^{}[ \mu \cos \beta+
\epsilon _iM_{\chi _i^0} \sin \beta ]}}\;, \nonumber \\ \\
{{z_{i4}}\over {z_{i1}}}&=&{{-\epsilon _iM_{\chi _i^0}
[M_2-\epsilon _iM_{\chi _i^0}]
[M_1-\epsilon _iM_{\chi _i^0}]-M_Z^2\cos ^2\beta
[(M_1-M_2)\cos ^2\theta _W^{}+M_2-\epsilon _iM_{\chi _i^0}]}\over
{M_Z[M_2-\epsilon _iM_{\chi _i^0}]\sin \theta _W^{}[ \mu \cos \beta+
\epsilon _iM_{\chi _i^0} \sin \beta ]}}\;, \nonumber \\
\end{eqnarray}
\end{mathletters}
and
\begin{eqnarray}
z_{i1}=\left [1+\left ({{z_{i2}}\over {z_{i1}}}\right )^2
+\left ({{z_{i3}}\over {z_{i1}}}\right )^2+\left ({{z_{i4}}
\over {z_{i1}}}\right )^2
\right ]^{-1/2}\;.
\end{eqnarray}

In most SUSY models, the lightest neutralino ($\chi^0_1$) is
the preferred the dark matter particle;
it is a mixture of gauginos and higgsinos,
\begin{equation}
\chi^0_1 = z_{11}\tilde{B}     +z_{12} \tilde{W}^3
          +z_{13}\tilde{H}^0_1 +z_{14} \tilde{H}^0_2.
\end{equation}
The gaugino fraction of $\chi^0_1$ can be defined as
\begin{equation}
f_G = z_{11}^2 +z_{12}^2
\label{eq:fG}
\end{equation}
and its higgisino fraction as
\begin{equation}
f_H = z_{13}^2 +z_{14}^2.
\label{eq:fH}
\end{equation}
The lightest neutralino is more gaugino-like for $\mu > M_2$
and gauginos annihilate mostly into heavy quarks;
the lightest neutralino becomes more higgisino-like for $\mu \alt M_2$
and the higgsinos annihilate dominantly into gauge bosons.

The minimal supersymmetric model has 63 free parameters
with real mass matrices and couplings.
To make the analysis of the MSSM tractable,
we assume a common value for the masses of scalar fermions
and the trilinear couplings ($M_{\susy} = m_{\tilde{f}} = A_f$).
The most relevant of the remaining free parameters are
the SU(2) gaugino mass ($M_2$), the Higgs mixing parameter ($\mu$),
the ratio of vacuum expectation values (VEVs) of Higgs fields
($\tan\beta \equiv v_2/v_1$) and the CP-odd Higgs-boson mass ($m_A$).

\subsection{The Effects of Muon Energy Threshold}

The indirect search for SUSY neutralino dark matter 
with high energy neutrinos is a promising 
approach to detect the annihilations of neutralino dark matter 
accumulated in the Sun \cite{Silk}-\cite{Feng01}. 
To obtain realistic rate estimates, we consider 
the muon detection threshold, quark hadronization,
and solar absorption.

For neutrinos from $\chi\bar{\chi} \to W^+ W^- \to \mu^+ \nu_\mu +X$
the average energy of the muon is about $m_\chi/4$; it is about $m_\chi/6$
for neutrinos from
$\chi\bar{\chi} \to Q \bar{Q} \to \mu^+ \nu_\mu q +X$.
The detection energy threshold for muons makes a significant impact on the  
indirect detection rate.
The effect of muon threshold on the indirect detection rate
from MSSM neutralino dark matter annihilations in the Sun 
is demonstrated in Figs. 1, 2 and 3 
for $m_{\chi^0_1} = 200 \; {\gev}$, $500 \; {\gev}$, and $1000 \;{\gev}$, 
with two values of $\tan\beta$ and two sets of $M_2$ and $\mu$: 
(i) $\mu \agt M_2$ such that the lightest neutralino is more gaugino-like, and 
(ii) $\mu < M_2$ such that the $\chi^0_1$ is more higgsino-like.
For simplicity, $M_{\rm SUSY} = 1.5 m_{\chi^0_1}$ is assumed for
a common squark mass ($m_{\tilde{q}}$),
a common slepton mass ($m_{\tilde{\ell}}$),
and a common trilinear coupling ($A_u = A_d = A_\ell$).
We note that the higgsino-like neutralino has a substantially higher rate
with a higher average muon energy than a gaugino-like neutralino.
In addition, a higher detector energy threshold effectively reduces 
the indirect detection rate for background and 
the signal with $m_{\chi^0_1} \alt 200$ GeV.
For $m_{\chi^0_1} \agt 500$ GeV, the indirect detection rate is 
not very sensitive to the muon energy threshold.

\subsection{Hadronization of Quarks and Solar Absorption}

There are uncertainties in the signal rate calculation
form hadronization of quarks and from solar absorption
that we now discuss.
For a gaugino-like neutralino, the dominant annihilation modes are
$b\bar{b}$ or $\tau \bar{ \tau }$ for $m_{\chi}<m_{W^{\pm}}$ and
$b\bar{b}$ for $m_{\chi}>m_{W^{\pm}}$.
For a higgsino-like neutralino, the dominant modes are $Z Z$ or $W^{+}W^{-}$
for $m_{\chi}<m_{t}$ and $t\bar{t}$ for $m_{\chi}>m_{t}$ \cite{Jungman1995}.
The effects of hadronization of heavy quarks into heavy mesons or heavy  
baryons reduces the energy of the neutrinos from the semileptonic  
decays\cite{Jungman1995}.

Moreover, neutrinos produced in the Sun may be absorbed
before they escape the solar medium.
We use the Ritz and Seckel parameterization up to energies around
1 TeV for all neutrino producing annihilation modes \cite{Ritz} to estimate  
the neutrino absorption losses.
Above the TeV scale, extra corrections are treated in a manner similar
to work by Edsjo \cite{Edsjo}.

Indirect detectors generally have neutrino energy thresholds
of at least a few tens of GeV.
We have chosen 25 GeV for the muon energy threshold in our analysis
which corresponds to a neutrino energy of about 50 GeV.

The combined effects of hadronization, solar absorption
and muon threshold energy can vary from total obliteration of the neutrino signal
for a neutralino mass around 100 GeV to merely a reduction by less than one half 
in neutrino signal for a neutralino mass close to 1 TeV.
Figures 1, 2 and 3 show representative calculations for a typical MSSM neutralino 
dark matter with a mass of 200 GeV, 500 GeV and 1000 GeV, respectively.
The effects become far less significant for heavier neutralinos.
However, higher energy neutrinos have a larger solar absorption rate,
therefore, slightly more neutrinos from annihilations of 1000 GeV neutralinos are
absorbed than those from 500 GeV neutralinos.

\subsection{Relic Density and Indirect Detection Rate}

For large $\tan\beta$ with heavy SUSY particles, 
the neutralino annihilation cross section is enhanced 
by the s-channel diagrams involving broad resonance poles 
of Higgs bosons,  
$\chi^0_1 \chi^0_1 \to A^0,H^0 \to b\bar{b}, \tau\bar{\tau}$, 
where $A^0$ is the CP odd pseudoscalar ($A$), 
and $H^0$ is the heavier CP even scalar ($H$).
Consequently, the neutralino relic density lies within 
the cosmologically interesting region
in much of the parameter space in SUSY models 
\cite{relic1,relic2,relic3,omhs,relic5,relic6}.

In Figure 4,
we present relic density of the neutralino dark matter ($\Omega_{\chi^0_1} h^2$)
versus $M_2$ for several values of $\mu$ with
$M_{\susy}$ taken to be the larger of 300 GeV and 1.5$m_{\chi^0_1}$.
For 200 GeV $\alt M_2 \alt$ 1200 GeV with 3 $\alt \tan\beta \alt$ 50
and $M_{\susy} \simeq {\rm max}(300 \; \gev, 1.5m_{\chi^0_1})$.
the relic density can be in the cosmological interesting region.

The indirect detection rate ($dN_{\id}/dA$) in events/$\km^2$/year
is presented versus $M_2$ in Fig.~5.
Here we have assumed a detector threshold energy of muons to be 25 GeV.
Figure 6 shows regions in the parameter plane of ($M_2,\mu$)
that may yield an indirect detection rate in events/$\km^2$/year:
(i) $dN_{\id} > 100$ (dark shading), 
(ii) $100 > dN_{\id} > 10$ (intermediate)
(iii) $dN_{\id} < 10$ (light shading).
Parameters in the blank regions do not generate a cosmologically interesting 
relic density ($0.05 \leq \Omega_{\chi^0_1} \leq 0.3$) 
for the neutralino dark matter.

The indirect detection rate can be significant
($> 10$ events/$\km^2$/year) in large regions
of the parameter space with $10 \alt \tan\beta \alt 50$:
(i) in the neighborhood of $M_2 \sim$ 500 GeV and $\mu \sim 500$ GeV,
(ii) in the neighborhood of $M_2 \sim$ 4000 GeV and $\mu \sim 1200$ GeV,
(iii) in a narrow band with $M_2 \agt 1200$ GeV and $M_2 \sim 2\mu$, and
(iv) in a narrow band with $M_2 \sim 400$ GeV and $\mu \agt 1200$ GeV
for $\tan\beta \sim 50$.

\section{The Minimal Supergravity Model}

\vspace{0.2in}

In the minimal supergravity (mSUGRA) model
it is assumed that SUSY is broken in a hidden sector
with SUSY breaking communicated to the observable sector
through gravitational interactions,
leading naturally to
a common scalar mass ($m_0$),
a common gaugino mass ($m_{1/2}$),
a common trilinear coupling ($A_0$),
and a bilinear coupling ($B_0$),
at the GUT scale ($M_{\rm GUT} \sim 2 \times 10^{16}$ GeV).
Through minimization of the Higgs potential, the $B$ parameter
and magnitude of the superpotential Higgs mixing parameter $\mu$
are related to $\tan\beta$ and $M_Z$.

The mass matrix of the charginos in the basis of the weak eigenstates
($\tilde{W}^\pm$, $\tilde{H}^\pm$) has the following form
\begin{equation}
M_C=\left( \begin{array}{c@{\quad}c}
M_2 & \sqrt{2}M_W\sin \beta \\
\sqrt{2}M_W\cos \beta & \mu
\end{array} \right)\;.
\label{eq:xino}
\end{equation}
The sign of the $\mu$ term in Eq.~(\ref{eq:xino})
establishes our sign convention. 
Analyses of $b\to s\gamma$ decay 
\cite{CLEO,LEP2,bsg1,bsg2,bsg3,bsg4,bsg4}
and muon g-2 measurements \cite{E821}-\cite{muon19}
strongly favor $\mu > 0$. 
We shall only consider $\mu >0$ and $A_0=0$. 

The SUSY particle masses and couplings at $M_Z$ can be predicted
by the evolution of renormalization group equations (RGEs)
from $M_{\rm GUT}$.
The gaugino masses at the weak scale ($M_Z$)
have the following relations:
\begin{itemize}
\item $M_1/\alpha_1 = M_2/\alpha_2 = M_3/\alpha_3$,
\item $M_1 \simeq 0.4 m_{1/2}, \;\; M_2 \simeq 0.8 m_{1/2}, \;\;
 M_3 \simeq 2.7 m_{1/2}$,
\item $m_{\chi^0_1} : m_{\chi^0_2} : m_{\chi^\pm_1} : m_{\tilde{g}}
\simeq 1: 2: 2 :6$.
\end{itemize}
We calculate SUSY particle masses at $M_Z$
and the masses and couplings in the Higgs sector
with one loop corrections in the one-loop effective potential
at the scale
$Q = \sqrt{m_{\tilde{t}_L}m_{\tilde{t}_R}}$ \cite{Baer97,omhs}.
At the scale $Q$, the RGE improved one-loop corrections approximately
reproduce the dominant two loop corrections
\cite{twoloop1,twoloop2,twoloop3,twoloop4,twoloop5,twoloop6}
to the mass of the lighter CP-even scalar.

We impose the following theoretical requirements on the RGE evolution:
\begin{itemize}
\item[(i)] radiative electroweak symmetry breaking (EWSB) is achieved,
\item[(ii)] the correct vacuum for EWSB is obtained (tachyon free), and
\item[(iii)] the lightest SUSY particle is the lightest neutralino.
\end{itemize}

In Figure 7, we present the relic density of
the mSUGRA neutralino dark matter ($\Omega_{\chi^0_1} h^2$)
versus $m_{1/2}$ for several values of $m_0$.
For $\tan\beta \alt 35$, 
cosmologically interesting relic densities are found
for $m_{1/2} \alt 500$ GeV.
For $\tan\beta \alt 50$, dips in $\Omega_{\chi^0_1}$ appear near
resonance poles at $2 m_{\chi^0_1} \simeq m_H,m_A$.

Figure 8 shows the higgsino fraction of the lightest neutralino
in the mSUGRA model. In most the of the $m_{1/2},m_0$ plane,
$\chi^0_1$ is mostly gaugino-like even with $m_0 > 1000$ GeV
\cite{Toby,FMW}.
Only in a narrow band when values of $m_{1/2}$ and $m_0$ are
close to theoretically excluded region,
can the lightest neutralino have a higgsino fraction larger than 10\%.
The region excluded by the $m_{\chi^+_1} \alt 103$ GeV limit
from the chargino search \cite{LEP2} at LEP2 is indicated.       

The indirect detection rate ($dN_{\id}/dA$) in events/$\km^2$/year
versus $m_{1/2}$ is presented in Fig. 9 for the mSUGRA model.
Again we take the detector threshold energy of the muon to be 25 GeV.
Figure 10 shows regions in the parameter plane of ($m_{1/2},m_0$)
that may yield an indirect detection rate in events/$\km^2$/year with:
(i) $10 > dN_{\id}/dA > 1$ (intermediate-shading) and 
(ii) $dN_{\id}/dA < 1$ (light shading).
There is no region that has $dN_{\id}/dA > 10$.
Parameters in the blank regions do not generate a cosmologically interesting 
relic density ($0.05 \leq \Omega_{\chi^0_1} \leq 0.3$) 
for the neutralino dark matter.
For $\tan\beta =50$, the relic density is suppressed below the
cosmologically interesting value when $2m_{\chi^0_1} \simeq m_H,m_A$.
That leads to a blank narrow band in Fig. 10(b).

\section{Supergravity Models with Nonuniversal Higgs-Boson Masses
at the GUT Scale}

We next consider a supergravity (SUGRA) unified model
with non-universal boundary conditions for the Higgs bosons
at the unified scale ($m_{\rm GUT}$).
We parameterize the GUT-scale Higgs masses as\footnote{
In Refs. \cite{nonuniversal1,nonuniversal2},
$m_{H_i}^2 ({\rm GUT})$ are parameterized to be $(1+\delta_i) m_0^2$.}
\begin{eqnarray}
m_{H_i} ({\rm GUT}) = (1+\delta_i) m_0 = \rho_i m_0, \; i = 1,2
\end{eqnarray}
where $\delta_i = -1$ and 2 correspond to $\rho_i = 0$ and $2$
The universal mSUGRA model has $\delta_1 = \delta_2 =0$,
while $\delta_i = -1$ and $\delta_i = 2$ correspond to
$m_{H_i}({\rm GUT}) = 0$ and $m_{H_i}({\rm GUT}) = 2m_0$.
The nonuniversality of Higgs-boson masses at $m_{\rm GUT}$
can significantly affect the values of Higgs masses and couplings
at the weak scale.

In Figures 11 and 12 we present the relic density of neutralino dark matter
($\Omega_{\chi^0_1} h^2$)
versus $\delta_1$ and $\delta_2$ for several values of $m_0$.
For $\tan\beta \alt 10$,
the non-universality of the Higgs masses makes slight only impact on
relic density of the neutralino dark matter.
For $\tan\beta \agt 35$,
non-universality affects the weak-scale Higgs-bosons masses,
the value of the $\mu$ parameter and
the neutralino relic density.
For $\tan\beta \sim 35$ and 50,
dips appear when $2m_{\chi^0_1}$ becomes close to $m_H,m_A$.

In Fig. 13, we present the higgsino fraction of 
the lightest neutralino ($f_H$) as defined in Eq.~(\ref{eq:fH})
in the plane of ($\delta_1,\delta_2$)
with $m_{1/2} = m_0 =$ 400 GeV, $A_0 = 0$.
For $\tan\beta \sim 3$, an increase in $\delta_1$ or $\delta_2$
generates a larger Higgsino fraction for $\chi^0_1$,
while for $\tan\beta \agt 10$, a decrease in $\delta_1$ or
an increase in $\delta_2$ can make $f_H$ larger for $\chi^0_1$.
The higgsino fraction of $\chi^0_1$ is very sensitive to
the value of $\delta_2$.

The indirect detection rate ($dN_{\id}/dA$) in events/$\km^2$/year
versus $m_{1/2}$ is presented in Figs. 14 and 15.
The threshold detection energy of the muon is assumed be 25 GeV.
For $\tan\beta \agt 35$, the indirect detection rate can be
large for chosen values of the parameters.
Figures 16, 17 and 18 show regions
in the parameter plane of ($m_{1/2},m_0$)
that may yield an indirect detection rate in events/$\km^2$/year,
(i) $dN_{\id} > 10$ (dark shading), 
(ii) $10 > dN_{\id} > 1$ (intermediate shading)
(iii) $dN_{\id} < 1$ (light shading).
Three cases are considered:
(a) $\delta_1 = -0.5$ and $\delta_2 = 0$ (Fig. 16),
(b) $\delta_1 = 0$ and $\delta_2 = 0.5$ (Fig. 17), and
(a) $\delta_1 = -0.5$ and $\delta_2 = 0.5$ (Fig. 18).
Parameters in the blank regions do not generate a cosmologically interesting 
relic density ($0.05 \leq \Omega_{\chi^0_1} \leq 0.3$) 
for the neutralino dark matter.
For $\tan\beta \sim 50$, non-universality can significantly enhance
the indirect detection rate for the neutralino dark matter.

For $\tan\beta \sim 10$,
the predicted indirect detection rate is interesting
only in a very small band near the theoretically excluded region
with $m_{1/2} \agt 600$ GeV and $m_0 \sim 3 m_{1/2}$.
For $\tan\beta \sim 50$,
the indirect detection rate is significant in a large region of
parameter space with $m_{1/2} \alt 1000$ GeV and $m_0 \alt 800$ GeV.

\section{Prospects for Indirect Detection}

For a very large volume second generation neutrino detector with superb angular  
resolution and large statistics for atmospheric neutrinos of order  
$10^4$ events per year for IceCube, this background can be reduced, 
in fact, essentially eliminated. 
Because the atmopheric neutrino background will be  
known with high precision, the number of events in the angular bin of  
the Sun can be determined by interpolation from adjacent bins. 
It should be possible to estimate the rate of atmospheric neutrinos 
to the level of a few events, and to achieve a background much smaller  
than the fluctuations on the events in the bin. 
In such a detector, the dominant background comes from neutrinos produced 
by cosmic ray interacting in the Sun's corona. 
This background is between 10 and 5 events per year
per square kilometer for muon thresholds between 25 and 100 GeV \cite{Bergstrom}.

The next generation high energy neutrino telescopes such as IceCube
and ANTARES with effective area on the order of 1 $\km^2$ may have the  
capability to make the first detections of supersymmetric
cold dark matter through the detection of neutrinos produced via 
annihilation of heavy neutralino in the Sun.
Considering muon thresholds ranging from 25 GeV to 100 GeV
(reflecting a variety of detectors and their variation in design),
our calculations show that a significant fraction of the MSSM parameter space 
with $\tan\beta \agt 10$ should produce
a muon event rate above backgrounds in such detectors.
The prospects for searching mSUGRA
parameter space are less optimistic than in the MSSM.  
If non-universal boundary conditions are considered, the prospects of detecting 
SUGRA neutralino dark matter with such detectors are increased.

We have focused on the flux of muons ($\phi_\mu$) generated by 
high-energy neutrinos (in units of 'events/km$^2$/year'). 
The event rate for a thin detector such as the AMANDA can be obtained 
from $\phi_\mu \times A_{\rm eff}$, where $A_{\rm eff}$ is the effective area.
For detectors of a volume $\agt 1 \, \km^3$ such as the IceCube, 
the thin detector approximation is valid for muon energies larger than 
a few hundred GeV.
For lower energy muons, the event rates will be higher than 
those obtained from thin detector approximation \cite{Bergstrom}.

The IceCube detector will have good angular and energy resolution 
for high energy muons \cite{icecube}. 
The measurement of muon angular distribution 
can be used to focus the search window around the direction from the Sun
and to reduce the background effectively.
The atmospheric neutrino background in a solar pixel will be determined accurately.
The significance of the neutralino signal at the IceCube
will be greatly enhanced with suitable background subtraction.

The question of naturalness and fine tuning arises for models
with heavy neutralinos.  Recently, it has been shown that
TeV scale values of $m_0$ can be natural \cite{FMM}.
The purpose of this paper does not lie with this issue and we put
forward our results only as a scan of the described parameter space
setting aside the subjective criterion of naturalness requirements.
However, because of naturalness, we do not consider 
the super-heavy models in which the lightest neutralino 
has mass of order several TeV or greater.

\section{Conclusions}

High energy muons produced by neutrinos from relic neutralino
annihilation in the Sun can lead to promising signals 
in the ice and the water detectors of high energy neutrinos.
We find that: 

\begin{enumerate}
\item[(i)] 
The effects of detector threshold, hadronization and solar absorption 
can be essential in the evaluation of indirect detection rates, especially 
for neutralinos with mass below a few hundred GeV.
\item[(ii)] 
In large portions of the MSSM parameter space with $\tan\beta \agt 10$, 
the indirect detection rate is predicted to be greater than ten events/$\km^2$/year.
However, the indirect search for neutralino annihilation may be difficult 
for $\tan\beta \alt 10$ or $m_\chi \alt 200$ GeV.
\item[(iii)] 
The indirect search for the mSUGRA neutralino dark matter will be challenging. 
Only several years of observation with a square kilometer detector 
will likely lead to a discovery. 
SUGRA models with non-universal boundary conditions and $\tan\beta \agt 35$ 
give more interesting rates. 
\item[(iv)] 
For large values of large $\tan\beta$ 
the neutralino annihilation cross section and the indirect detection rate
are enhanced by the s-channel diagrams involving Higgs bosons,  
$\chi^0_1 \chi^0_1 \to A^0,H^0 \to b\bar{b}, \tau\bar{\tau}$, 
where $A^0$ is the CP odd pseudoscalar, 
and $H^0$ is the heavier CP even scalar, 
and high energy neutrinos are produced in the $b$ and $\tau$ decays.
\item[(v)] 
For the AMANDA experiment, with a muon energy threshold of 25 GeV 
the background muon flux from atmospheric neutrinos in a pixel containing 
the neutralino signal is about 
$4.6 \; {\rm events}/(10^4\; {\rm m^2})/{\rm year}$.
This background flux is larger than the signal rate 
in most regions of the parameter space in SUSY models. 
To improve the significance of signal over background, 
a higher muon energy threshold will be advantageous if $m_\chi \agt 500$ GeV.
\item[(vi)] 
For the IceCube and the ANTARES experiments, 
the large number of background events will make background subtraction possible 
with only a small remaining irreducible background \cite{Bergstrom}
and allow measurement of the neutralino annihilation signal.  
\item[(vii)] 
SUSY neutralino dark matter with a mass larger than about 200 GeV 
offers great promise for indirect detection experiments.
Together with direct detection experiments of neutralino dark matter 
and accelerator experiments at the upgraded Tevatron and 
the CERN Large Hadron Collider, high-energy neutrino telescopes 
will be able to survey large regions of parameter space beyond present experiments.
\end{enumerate}

\section*{Acknowledgments}

C.K. thanks Tzi-Hong Chieuh and George Hou for hospitality
at the National Taiwan University where part of the research was completed.
This research was supported in part by the U.S. Department of Energy
under Grants No. DE-FG02-95ER40896 and No. DE-FG03-98ER41066, 
and in part by the University of Wisconsin Research Committee
with funds granted by the Wisconsin Alumni Research Foundation.


\newpage


\begin{figure}
\centering\leavevmode
\epsfxsize=6in\epsffile{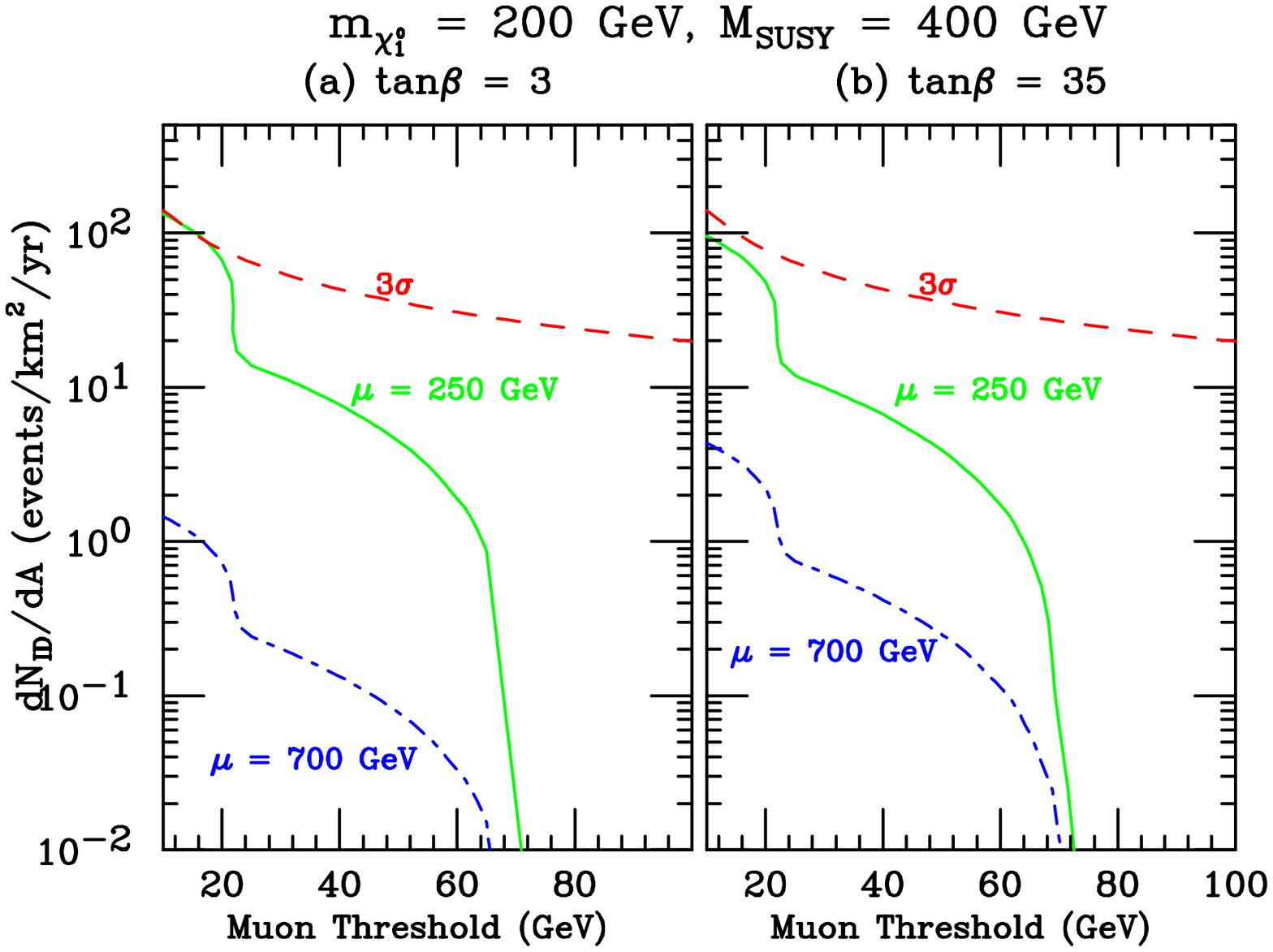}

\caption[]{
Indirect detection rate $dN_{\id}/dA$ in events/km$^2$/year
versus the muon detection energy threshold
for $m_{\chi_1^0}= 200$ GeV and $M_{\rm SUSY}= 400$ GeV
with $m_A = 175$ GeV and  
(a) $\tan\beta = 3$, $\mu = 700$ GeV, $M_2 = 400$ GeV,
and $\mu = 250$ GeV, $M_2 = 485$ GeV, as well as 
(b) $\tan\beta = 35$, $\mu = 700$ GeV, $M_2 = 400$ GeV,
 and $\mu = 250$ GeV, $M_2 = 445$ GeV. 
Also shown is the 3$\sigma$ signal rate ($=3\sqrt{N_B}$), 
where $N_B$ is the number of background events from the atmospheric neutrinos.
\label{fig:muon1}
}\end{figure}

\begin{figure}
\centering\leavevmode
\epsfxsize=6in\epsffile{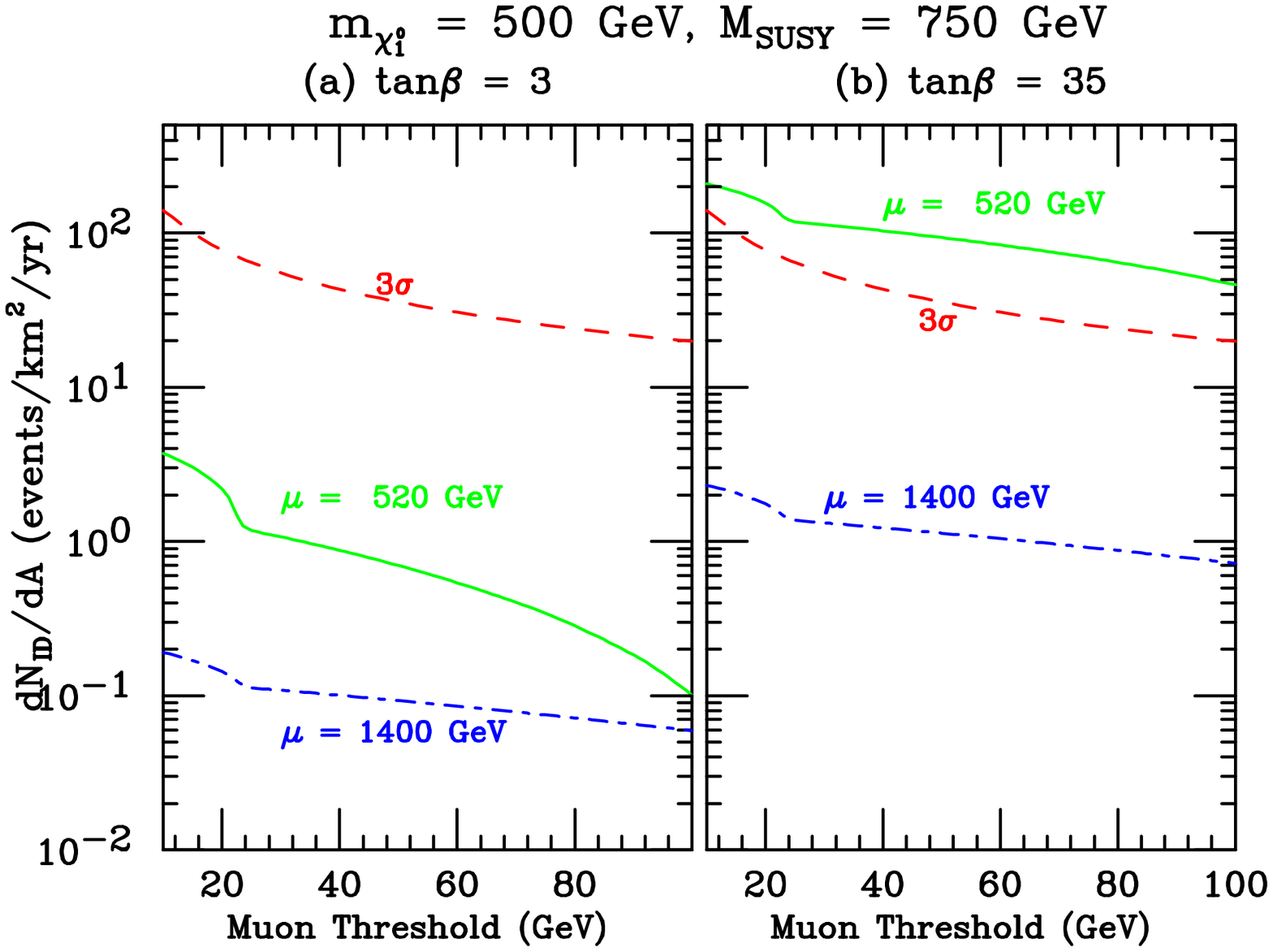}

\caption[]{
Indirect detection rate $dN_{\id}/dA$ in events/km$^2$/year
versus the muon detection energy threshold
for $m_{\chi_1^0}= 500$ GeV and $M_{\rm SUSY}= 750$ GeV
with $m_A = 500$ GeV and  
(a) $\tan\beta = 3$, $\mu = 1400$ GeV, $M_2 = 1000$ GeV,
and $\mu = 520$ GeV, $M_2 = 1280$ GeV, as well as 
(b) $\tan\beta = 35$, $\mu = 1400$ GeV, $M_2 = 1000$ GeV,
 and $\mu = 520$ GeV, $M_2 = 1160$ GeV. 
Also shown is the 3$\sigma$ signal rate ($=3\sqrt{N_B}$), 
where $N_B$ is the number of background events from the atmospheric neutrinos.
\label{fig:muon2}
}\end{figure}

\begin{figure}
\centering\leavevmode
\epsfxsize=6in\epsffile{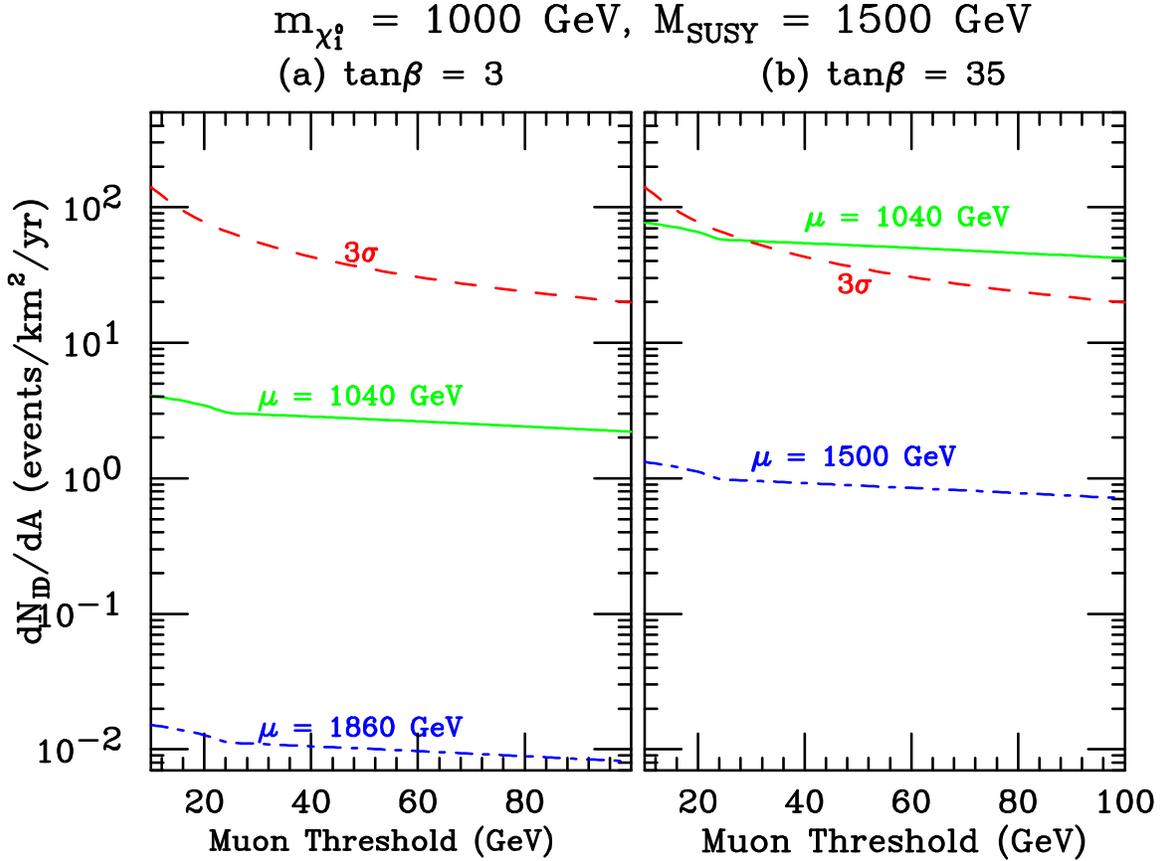}

\caption[]{
Indirect detection rate $dN_{\id}/dA$ in events/km$^2$/year
versus the muon detection energy threshold
for $m_{\chi_1^0}= 1000$ GeV and $M_{\rm SUSY}= 1500$ GeV
with $m_A = 1000$ GeV and  
(a) $\tan\beta = 3$, $\mu = 1040$ GeV, $M_2 = 2060$ GeV,
and $\mu = 1860$ GeV, $M_2 = 2000$ GeV, as well as 
(b) $\tan\beta = 35$, $\mu = 1040$ GeV, $M_2 = 2060$ GeV,
 and $\mu = 1500$ GeV, $M_2 = 2000$ GeV. 
Also shown is the 3$\sigma$ signal rate ($=3\sqrt{N_B}$), 
where $N_B$ is the number of background events from the atmospheric neutrinos.
\label{fig:muon3}
}\end{figure}


\begin{figure}
\centering\leavevmode
\epsfxsize=6in\epsffile{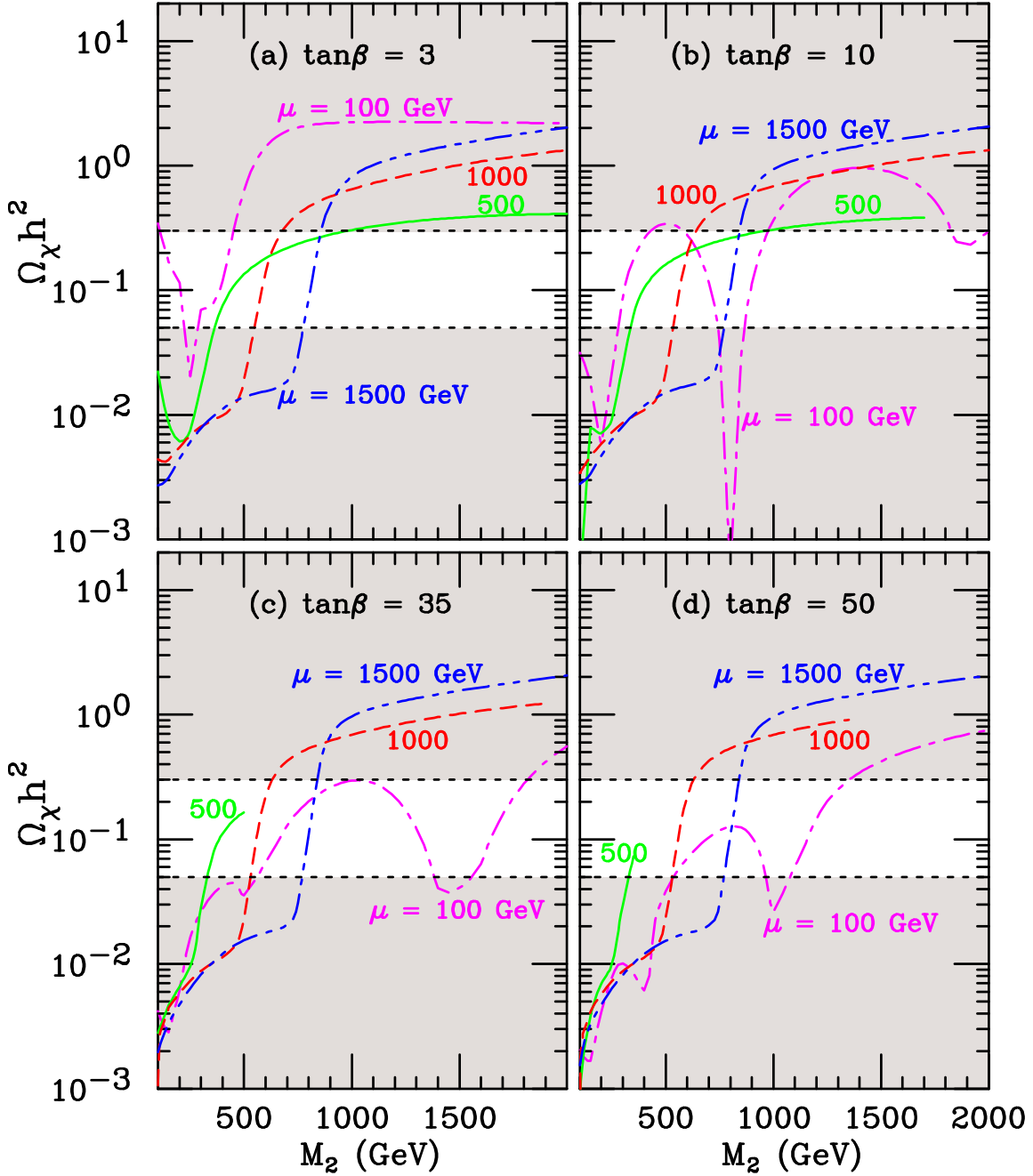}

\caption[]{
Relic density of the neutralino dark matter ($\Omega_{\chi^0_1} h^2$)
versus $M_2$ in the MSSM with several values of $\mu$
and $M_{\rm SUSY} = {\rm MAX}(300 \; {\gev}, 1.5 m_{\chi^0_1})$
for (a) $\tan\beta = 3$, (b) $\tan\beta = 10$, (c) $\tan\beta = 35$
and (d) $\tan\beta = 50$.
\label{fig:MSSM1}
}\end{figure}

\begin{figure}
\centering\leavevmode
\epsfxsize=6in\epsffile{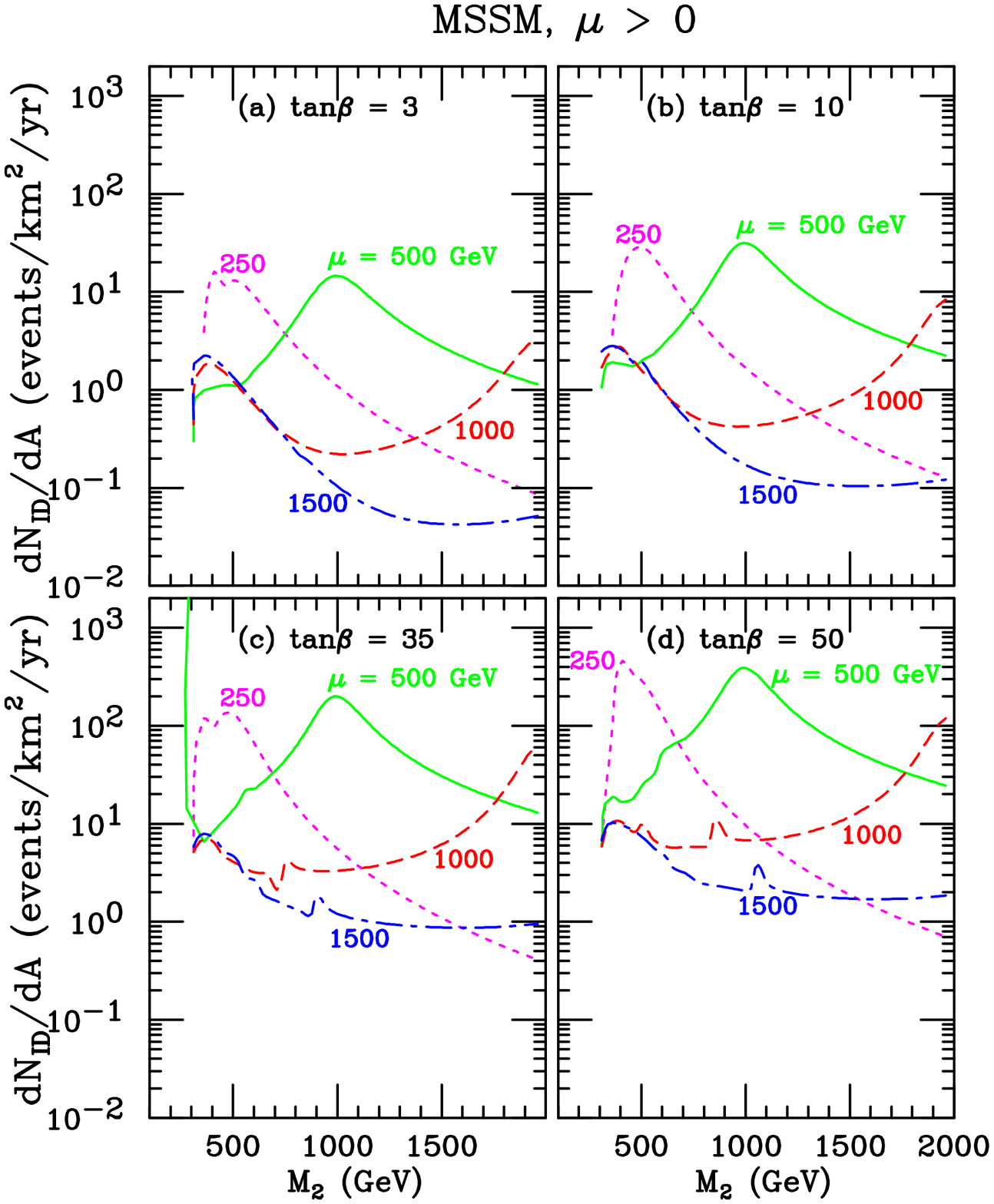}

\caption[]{
Indirect detection rate of the MSSM neutralino dark matter
from the Sun ($dN_{\rm ID}/dA$) in ${\rm Events}/\km^2/year$
versus $M_2$
with $M_{\rm SUSY} = {\rm MAX}(300 \; {\gev}, 1.5 m_{\chi^0_1})$
and several values of $\mu$
for (a) $\tan\beta = 3$, (b) $\tan\beta = 10$, (c) $\tan\beta = 35$
and (d) $\tan\beta = 50$.
\label{fig:MSSM2}
}\end{figure}
%

\begin{figure}
\centering\leavevmode
\epsfxsize=6in\epsffile{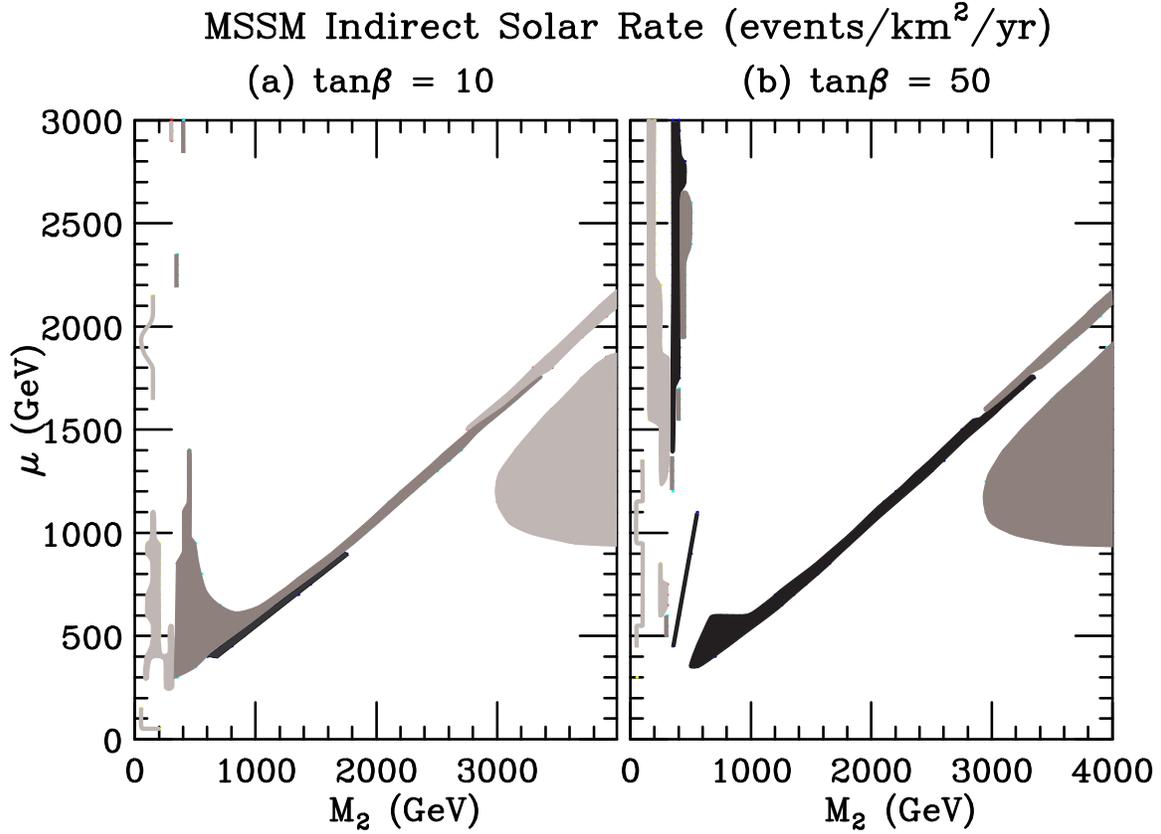}

\caption[]{
Regions of indirect detection rate 
of the MSSM neutralino dark matter in the ($M_2,\mu$) plane
with $M_{\rm SUSY} = {\rm MAX}(300 \; {\gev}, 1.5 m_{\chi^0_1})$
for (a) $\tan\beta = 10$ and (b) $\tan\beta = 50$.
The region with dark shading has $dN_{\id}/dA > 100$; 
the region with light shading has $dN_{\id}/dA < 10$; and 
the region with intermediate shading has $100 > dN_{\id}/dA > 10$ 
(event/km$^2$/year).
The blank regions do not have a cosmologically interesting relic density 
($0.05 \leq \Omega_{\chi^0_1} \leq 0.3$) for the neutralino dark matter.
\label{fig:MSSM3}
}\end{figure}
%


\begin{figure}
\centering\leavevmode
\epsfxsize=6in\epsffile{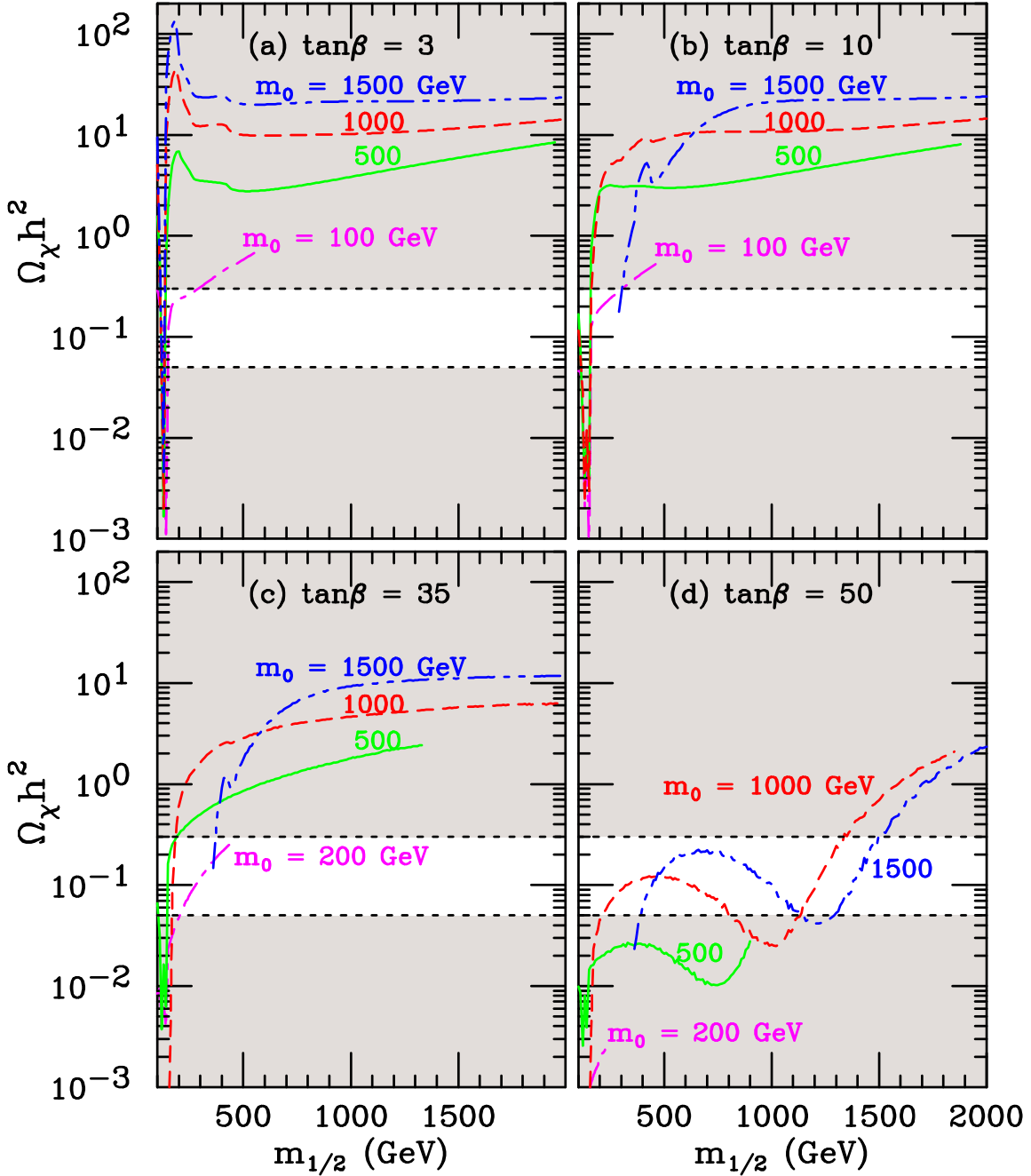}

\caption[]{
Relic density of the neutralino dark matter ($\Omega_{\chi^0_1} h^2$)
versus $m_{1/2}$ in the mSUGRA model
with $\mu >0$, $A_0 = 0$ and several values of $m_0$
for (a) $\tan\beta = 3$, (b) $\tan\beta = 10$, (c) $\tan\beta = 35$
and (d) $\tan\beta = 50$.
\label{fig:SUGRA1}
}\end{figure}

\begin{figure}
\centering\leavevmode
\epsfxsize=6in\epsffile{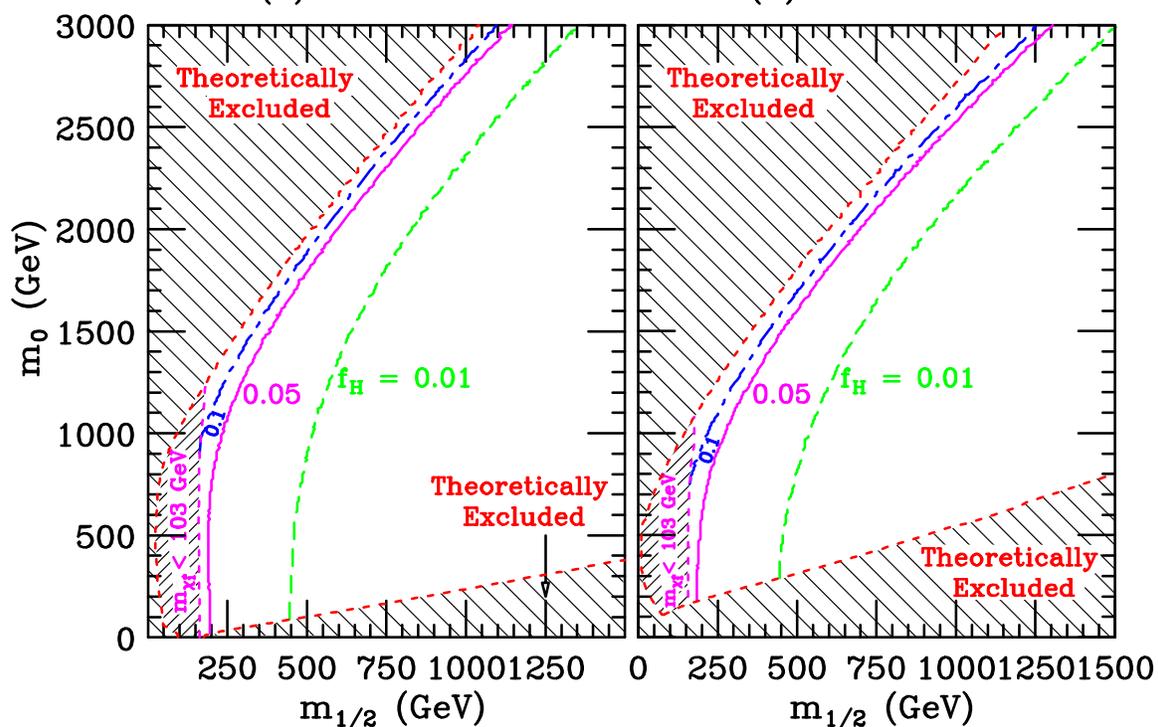}

\caption[]{
Contours of Higgsino fraction for the neutralino dark matter
of minimal supergravity in the ($m_{1/2},m_0$) plane
for (a) $\tan\beta = 10$ and (b) $\tan\beta = 50$.
Also shown are the parts of the parameter space
(i)~excluded by theoretical requirements,
or (ii)~excluded by the chargino search at LEP 2.
\label{fig:SUGRA2}
}\end{figure}
%

\begin{figure}
\centering\leavevmode
\epsfxsize=6in\epsffile{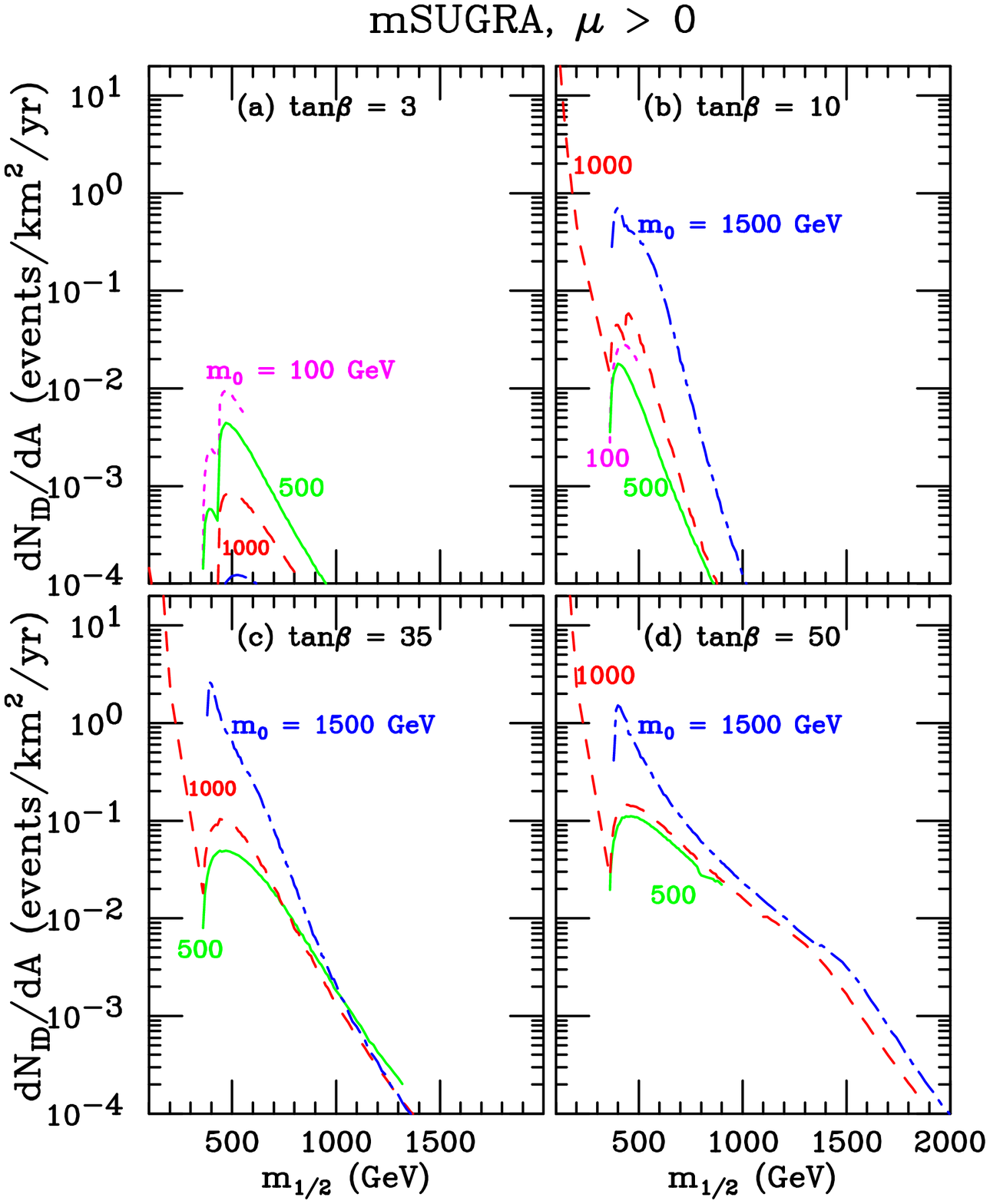}

\caption[]{
Indirect detection rate of the neutralino dark matter
from the Sun ($dN_{\rm ID}/dA$) in ${\rm Events}/km^2/year$
versus $m_{1/2}$
with $\mu >0$, $A_0 = 0$ and several values of $m_0$
for (a) $\tan\beta = 3$, (b) $\tan\beta = 10$, (c) $\tan\beta = 35$
and (d) $\tan\beta = 50$.
\label{fig:SUGRA3}
}\end{figure}
%

\begin{figure}
\centering\leavevmode
\epsfxsize=6in\epsffile{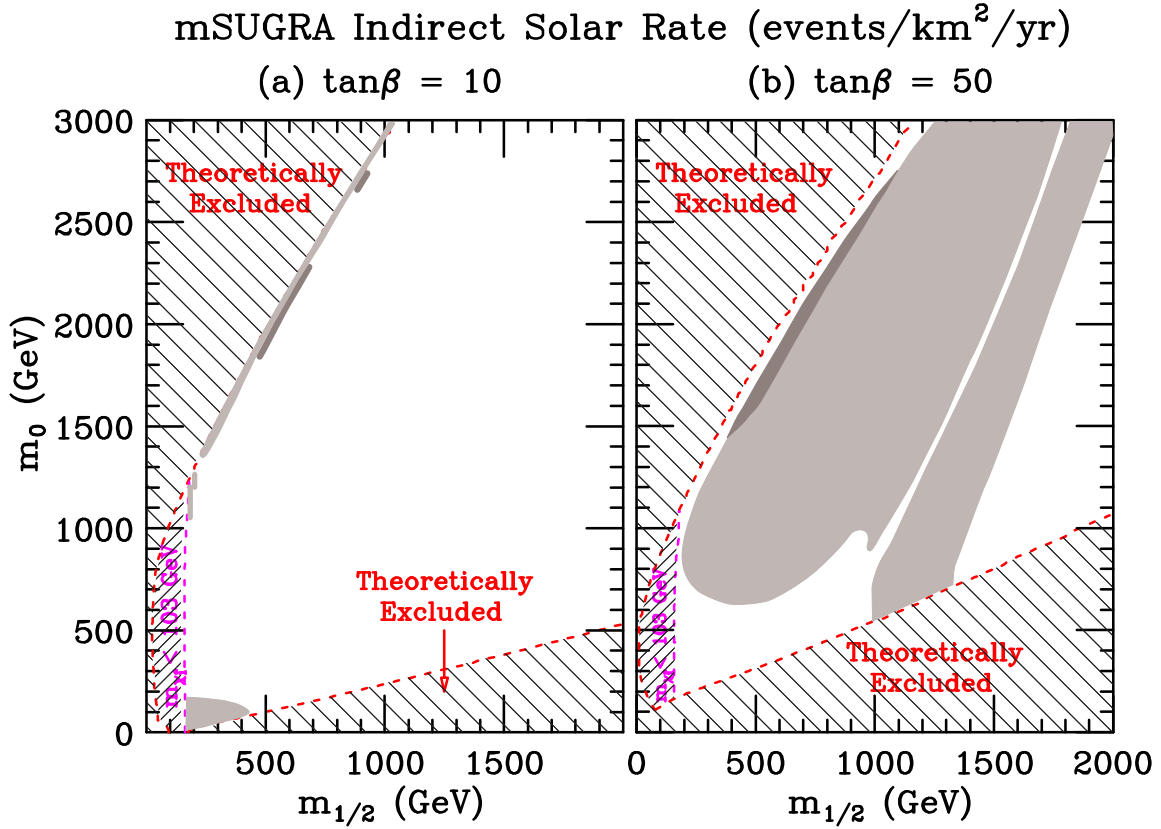}

\caption[]{
Regions of indirect detection rate
of the mSUGRA neutralino dark matter in the ($m_{1/2},m_0$) plane
with $\mu >0$ and $A_0 = 0$
for (a) $\tan\beta = 10$ and (b) $\tan\beta = 50$.
The region with light shading has $dN_{\id}/dA < 1$; and 
the region with intermediate shading has $10 > dN_{\id}/dA > 1$ (event/km$^2$/year).
There is no region with dark shading that has $dN_{\id}/dA > 10$.
The blank regions do not have a cosmologically interesting relic density 
($0.05 \leq \Omega_{\chi^0_1} \leq 0.3$) for the neutralino dark matter.
Also shown are the parts of the parameter space
(i)~excluded by theoretical requirements,
or (ii)~excluded by the chargino search at LEP 2.
\label{fig:SUGRA4}
}\end{figure}
%


\begin{figure}
\centering\leavevmode
\epsfxsize=6in\epsffile{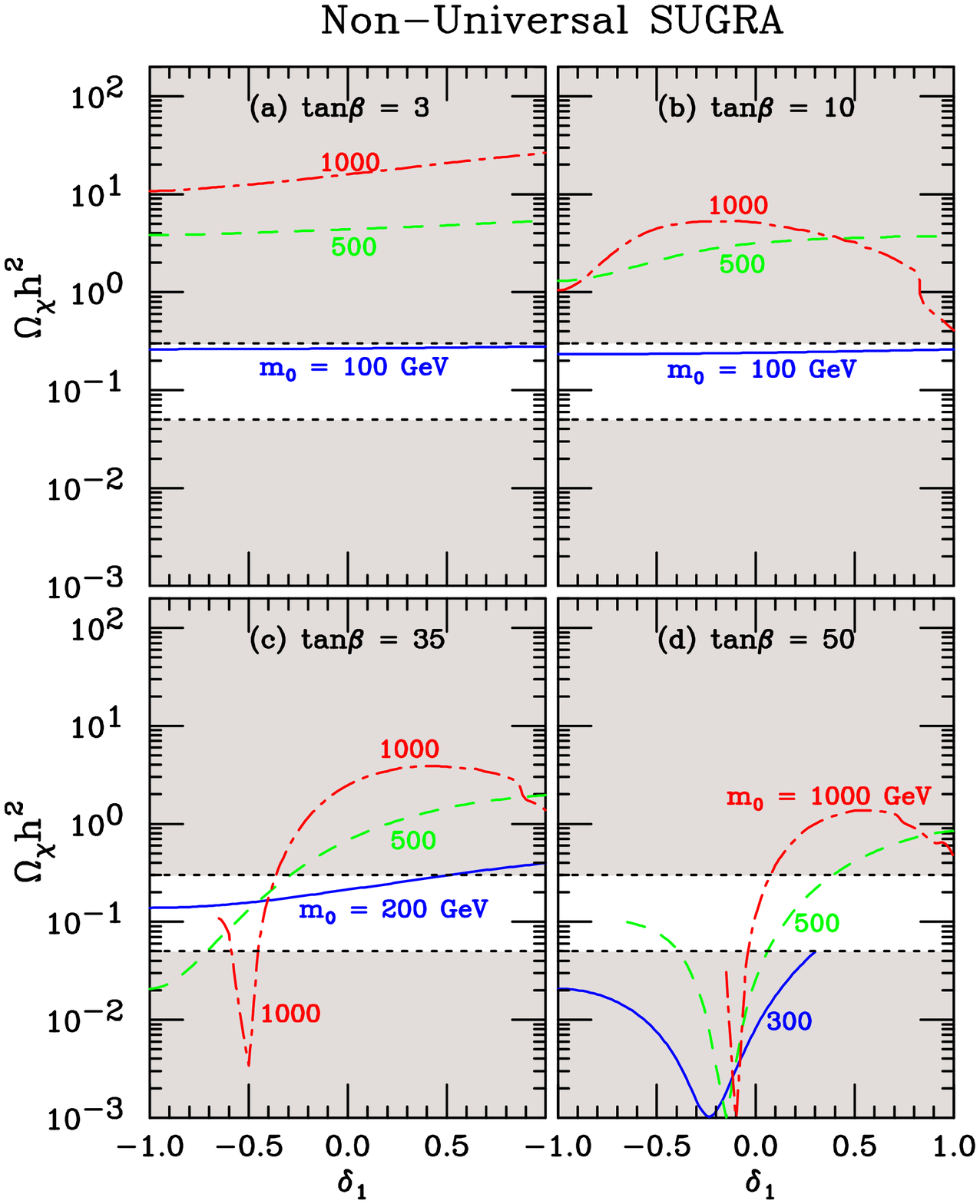}

\caption[]{
Relic density of the neutralino dark matter ($\Omega_{\chi^0_1} h^2$)
versus $\delta_1$
in a SUGRA model with non-universal Higgs masses at $M_{\rm GUT}$,
with $\delta_2 = 0$, $\mu >0$, $A_0 = 0$, and several values of $m_0$,
for $m_{1/2} = 250$ GeV with (a) $\tan\beta =  3$ and (b) $\tan\beta = 10$,
as well as
for $m_{1/2} = 400$ GeV with (c) $\tan\beta = 35$ and (d) $\tan\beta = 50$.
\label{fig:NONU1}
}\end{figure}

\begin{figure}
\centering\leavevmode
\epsfxsize=6in\epsffile{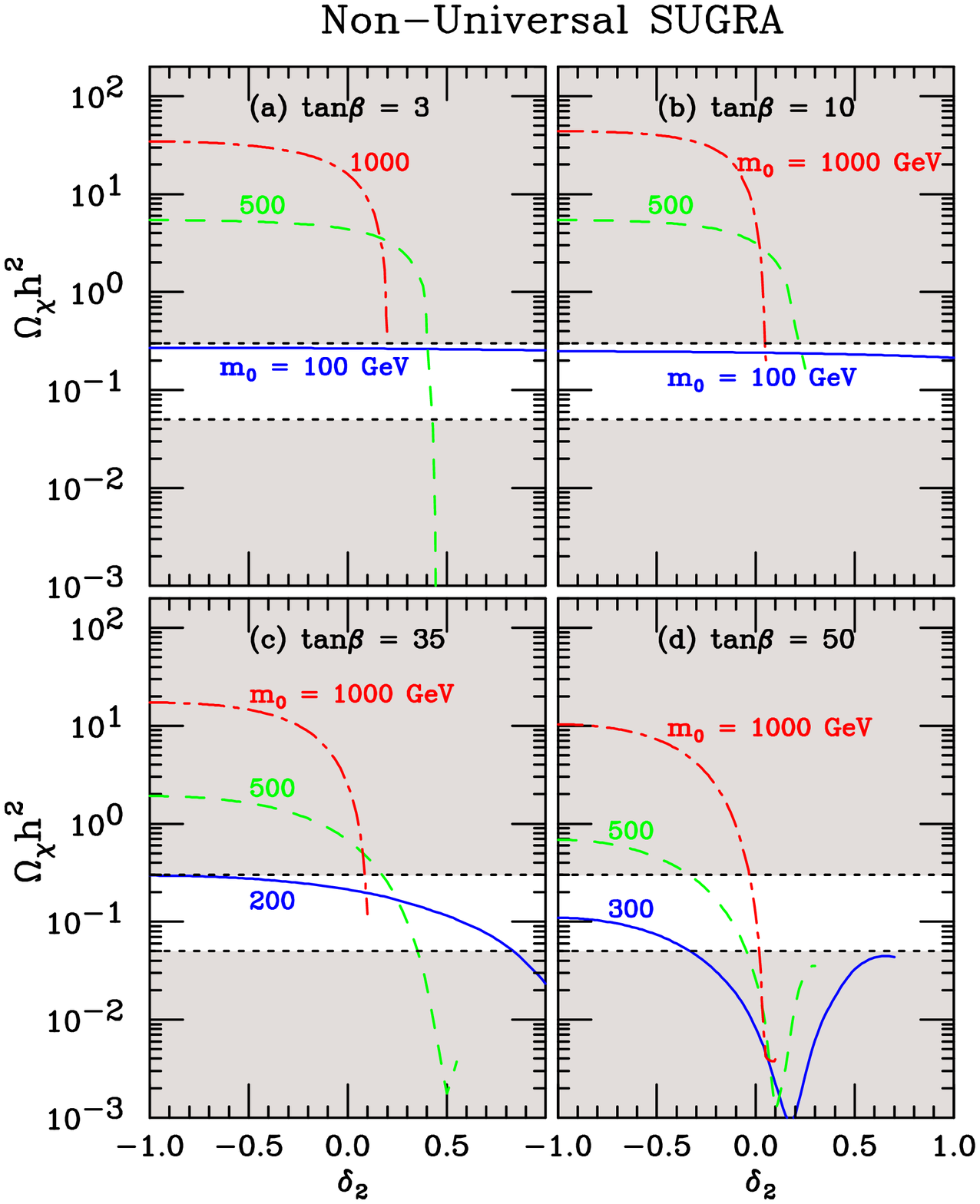}

\caption[]{
Relic density of the neutralino dark matter ($\Omega_{\chi^0_1} h^2$)
versus $\delta_2$
in a SUGRA model with non-universal Higgs masses at $M_{\rm GUT}$,
with $\delta_1 = 0$, $\mu >0$, $A_0 = 0$, and several values of $m_0$,
for $m_{1/2} = 250$ GeV with (a) $\tan\beta =  3$ and (b) $\tan\beta = 10$,
as well as
for $m_{1/2} = 400$ GeV with (c) $\tan\beta = 35$ and (d) $\tan\beta = 50$.
\label{fig:NONU2}
}\end{figure}

\begin{figure}
\centering\leavevmode
\epsfxsize=6in\epsffile{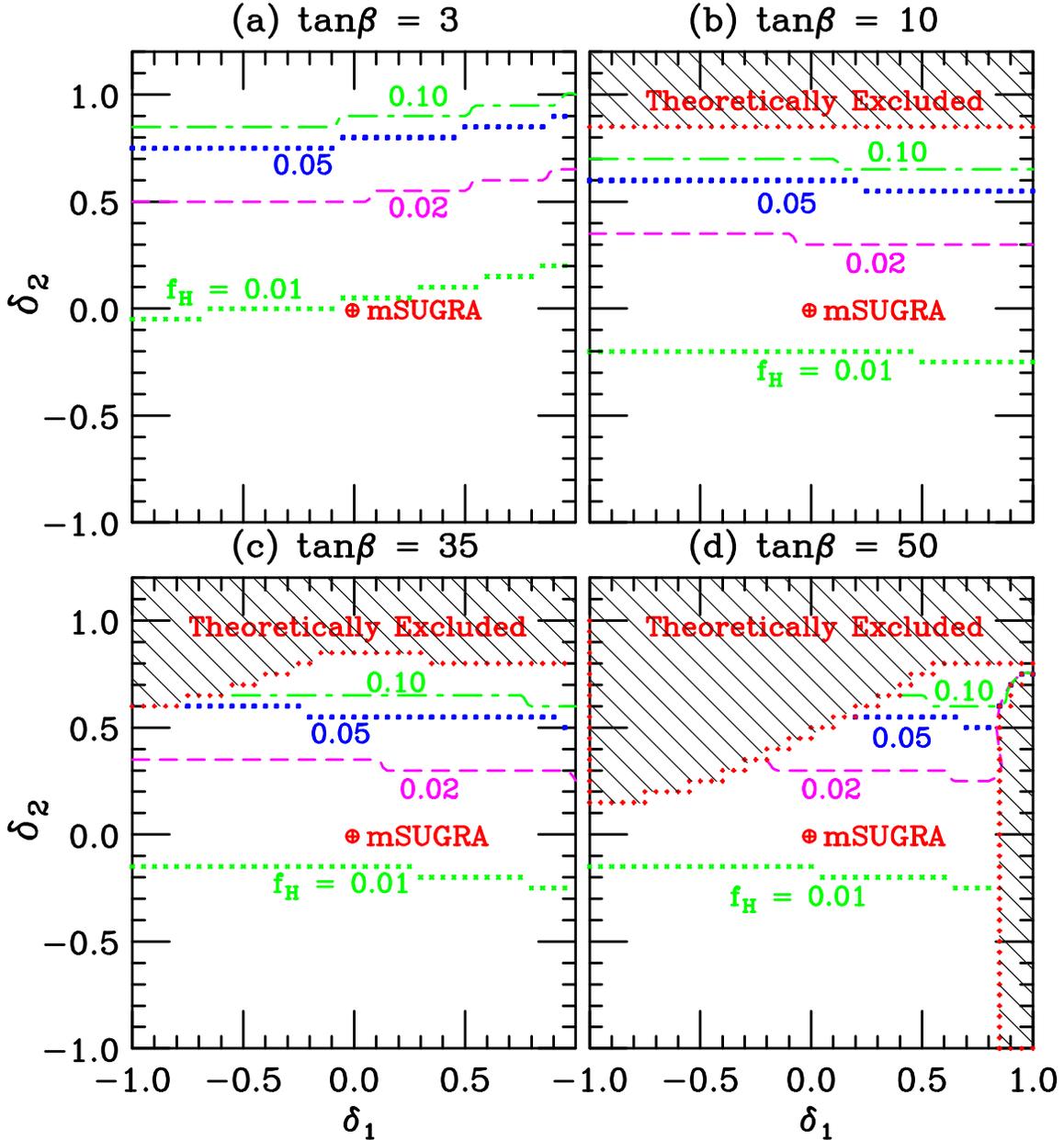}

\caption[]{
Contours of the Higgsino fraction of the neutralino dark matter ($f_H$)
in the ($\delta_1$,$\delta_2$) plan
within the framework of a sugravity unified model
with non-universal Higgs masses at $M_{\rm GUT}$,
$\mu >0$, $m_{1/2} = m_0 = 400$ GeV,
for (a) $\tan\beta = 3$, (b) $\tan\beta = 10$, (c) $\tan\beta = 35$
and (d) $\tan\beta = 50$.
\label{fig:NONU3}
}\end{figure}

\begin{figure}
\centering\leavevmode
\epsfxsize=6in\epsffile{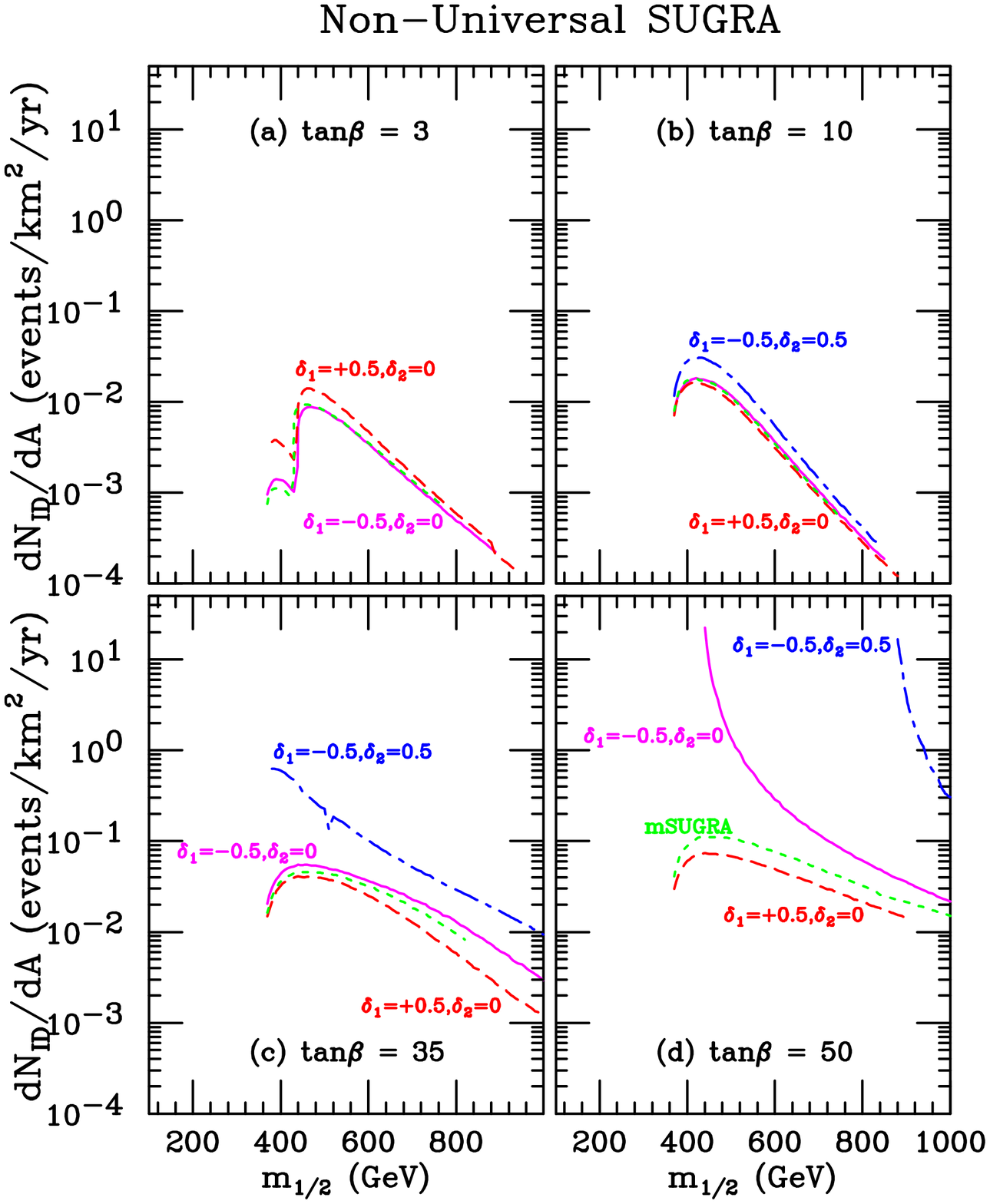}

\caption[]{
Indirect detection rate of the SUGRA neutralino dark matter
from the Sun in ${\rm Events}/\km^2/year$ versus $m_{1/2}$
with $\mu >0$, $A_0 = 0$,
and three choices of non-universal boundary conditions
for the Higgs masses at $M_{\rm GUT}$
(i)   $\delta_1 = -0.5$ and $\delta_2 = 0$,
(ii)  $\delta_1 = +0.5$ and $\delta_2 = 0$,
(iii) $\delta_1 = -0.5$ and $\delta_2 = +0.5$,
for $m_0 =  200$ GeV with
(a) $\tan\beta = 3$, (b) $\tan\beta = 10$ and (c) $\tan\beta = 35$,
as well as for $m_0 = 1000$ GeV with (d) $\tan\beta = 50$.
Also shown is the indirect detection rate in the mSUGRA
with $\delta_1 = \delta_2 = 0$.
\label{fig:NONU4}
}\end{figure}

\begin{figure}
\centering\leavevmode
\epsfxsize=6in\epsffile{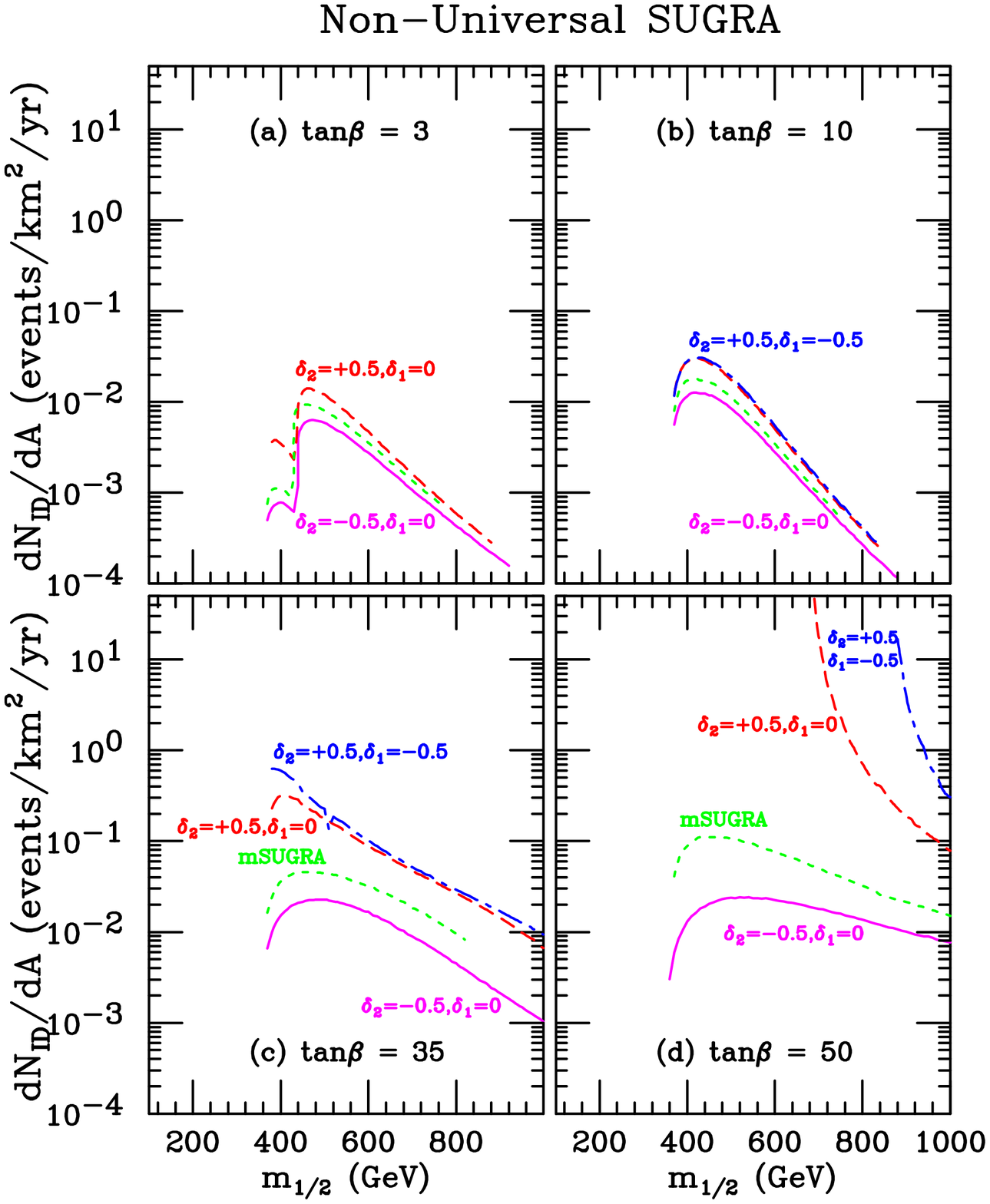}

\caption[]{
Indirect detection rate of the SUGRA neutralino dark matter
from the Sun in ${\rm Events}/\km^2/year$ versus $m_{1/2}$
with $\mu >0$, $A_0 = 0$,
and three choices of non-universal boundary conditions
for the Higgs masses at $M_{\rm GUT}$
(i)   $\delta_1 = 0$ and $\delta_2 = -0.5$,
(ii)  $\delta_1 = 0$ and $\delta_2 = +0.5$,
(iii) $\delta_1 = -0.5$ and $\delta_2 = +0.5$,
for $m_0 =  200$ GeV with
(a) $\tan\beta = 3$, (b) $\tan\beta = 10$ and (c) $\tan\beta = 35$,
as well as for $m_0 = 1000$ GeV with (d) $\tan\beta = 50$.
Also shown is the indirect detection rate in the mSUGRA
with $\delta_1 = \delta_2 = 0$.
\label{fig:NONU5}
}\end{figure}

\begin{figure}
\centering\leavevmode
\epsfxsize=6in\epsffile{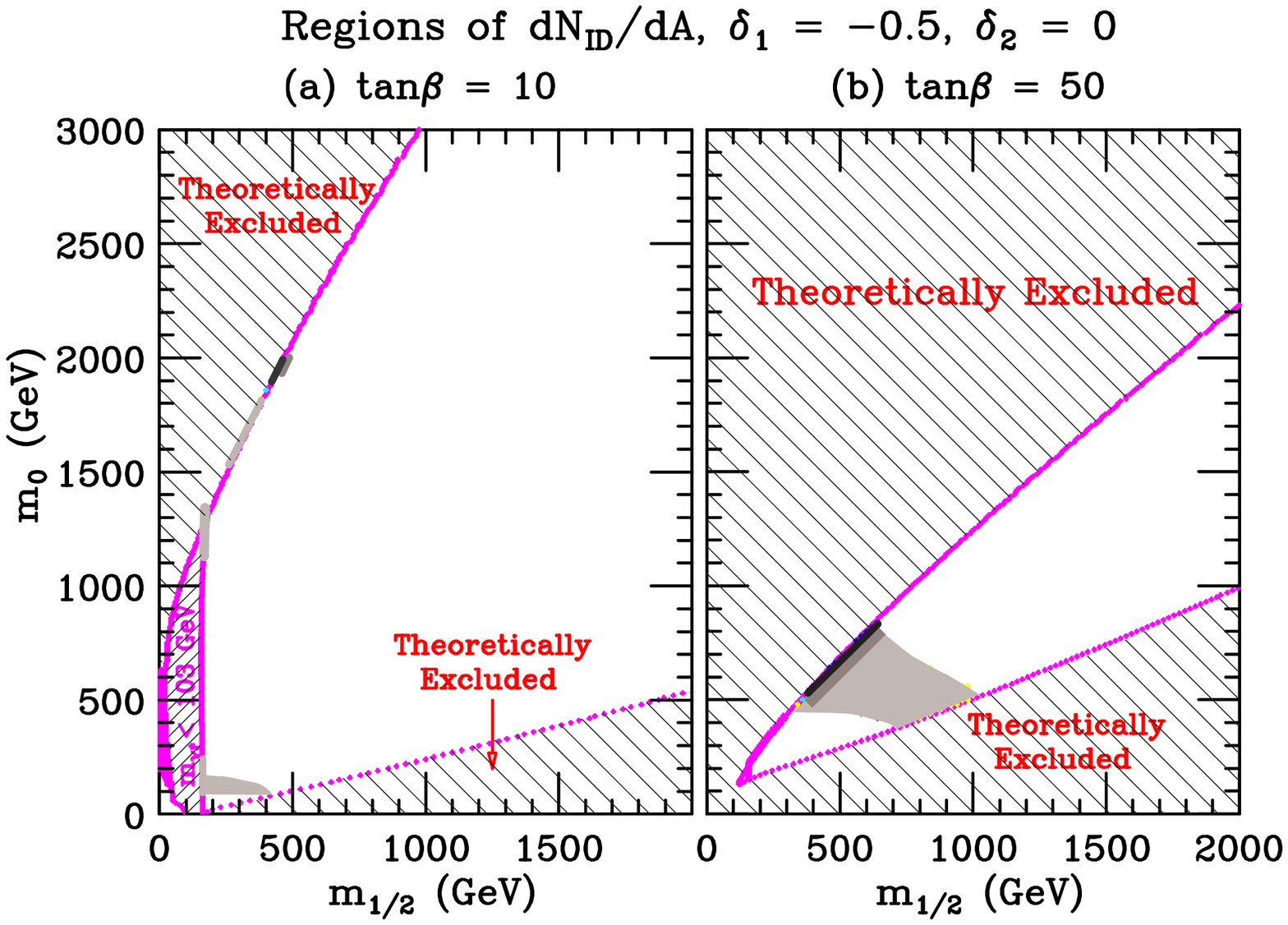}

\caption[]{
Regions of indirect detection rate of the neutralino dark matter
in the ($m_{1/2},m_0$) plane of a non-universal SUGRA model
with $\mu >0$, $A_0 = 0$ and non-universal boundary conditions
$\delta_1 =  -0.5$ and $\delta_2 = 0$,
for (a) $\tan\beta = 10$ and (b) $\tan\beta = 50$.
The region with dark shading has $dN_{\id}/dA > 10$; 
the region with light shading has $dN_{\id}/dA < 1$; and 
the region with intermediate shading has $10 > dN_{\id}/dA > 1$ (event/km$^2$/year).
The blank regions do not have a cosmologically interesting relic density 
($0.05 \leq \Omega_{\chi^0_1} \leq 0.3$) for the neutralino dark matter.
Also shown are the parts of the parameter space
(i)~excluded by theoretical requirements,
or (ii)~excluded by the chargino search at LEP 2.
\label{fig:NONU6}
}\end{figure}

\begin{figure}
\centering\leavevmode
\epsfxsize=6in\epsffile{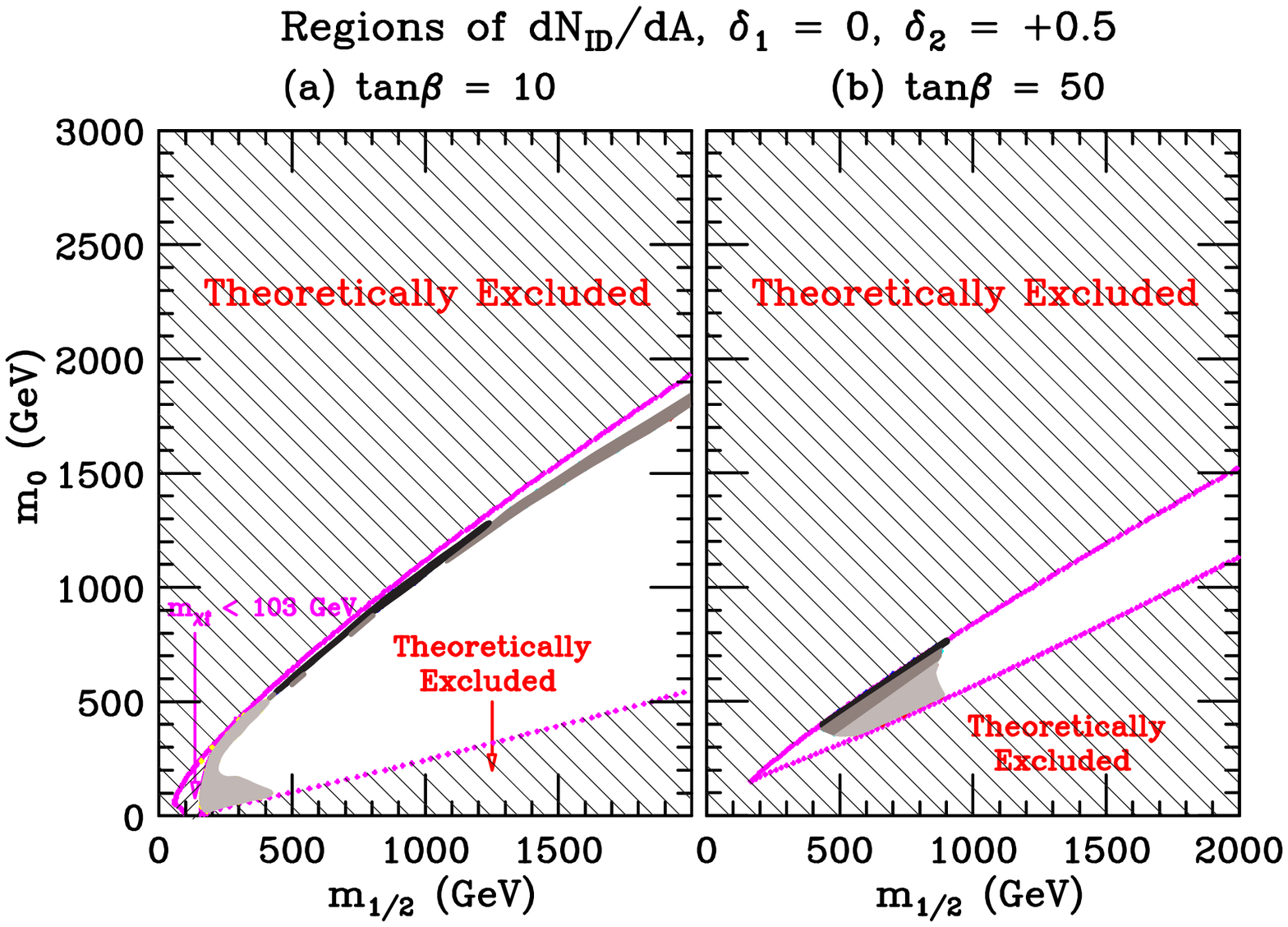}

\caption[]{
Regions of indirect detection rate of the neutralino dark matter
in the ($m_{1/2},m_0$) plane of a non-universal SUGRA model
with $\mu >0$, $A_0 = 0$ and non-universal boundary conditions
$\delta_1 =  0$ and $\delta_2 = +0.5$,
for (a) $\tan\beta = 10$ and (b) $\tan\beta = 50$.
The region with dark shading has $dN_{\id}/dA > 10$; 
the region with light shading has $dN_{\id}/dA < 1$; and 
the region with intermediate shading has $10 > dN_{\id}/dA > 1$ (event/km$^2$/year).
The blank regions do not have a cosmologically interesting relic density 
($0.05 \leq \Omega_{\chi^0_1} \leq 0.3$) for the neutralino dark matter.
Also shown are the parts of the parameter space
(i)~excluded by theoretical requirements,
or (ii)~excluded by the chargino search at LEP 2.
\label{fig:NONU7}
}\end{figure}

\begin{figure}
\centering\leavevmode
\epsfxsize=6in\epsffile{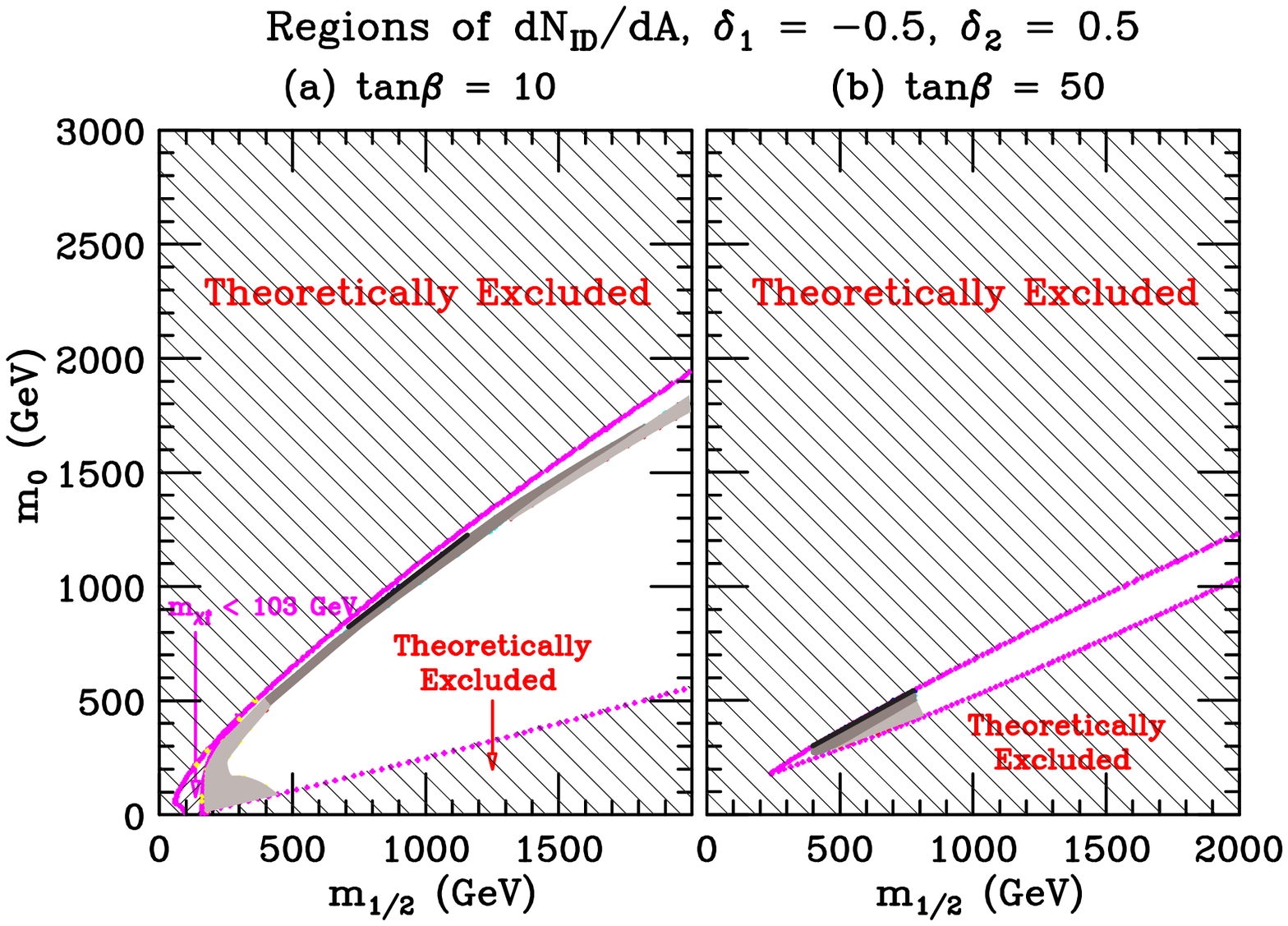}

\caption[]{
Regions of indirect detection rate of the neutralino dark matter
in the ($m_{1/2},m_0$) plane of a non-universal SUGRA model
with $\mu >0$, $A_0 = 0$ and non-universal boundary conditions
$\delta_1 =  -0.5$ and $\delta_2 = +0.5$,
for (a) $\tan\beta = 10$ and (b) $\tan\beta = 50$.
The region with dark shading has $dN_{\id}/dA > 10$; 
the region with light shading has $dN_{\id}/dA < 1$; and 
the region with intermediate shading has $10 > dN_{\id}/dA > 1$ (event/km$^2$/year).
The blank regions do not have a cosmologically interesting relic density 
($0.05 \leq \Omega_{\chi^0_1} \leq 0.3$) for the neutralino dark matter.
Also shown are the parts of the parameter space
(i)~excluded by theoretical requirements,
or (ii)~excluded by the chargino search at LEP 2.
\label{fig:NONU8}
}\end{figure}

\end{document}